\documentclass[superscriptaddress,amsmath,amssymb,aps,prd,tightenlines,nofootinbib,longbibliography]{revtex4-1}

\pdfoutput=1
\usepackage[utf8]{inputenc}
\usepackage[Symbolsmallscale]{upgreek}
\usepackage{tabularx}
\usepackage[protrusion=true,tracking=true,kerning=true,spacing=true,final,babel=true]{microtype}
\usepackage{xcolor}
\usepackage{graphicx}
\usepackage[final]{hyperref}
\usepackage{nth}
\usepackage{xcolor}

\usepackage{placeins}
\usepackage{epsfig}
\usepackage{hyperref}
\usepackage{calrsfs}
\DeclareMathAlphabet{\pazocal}{OMS}{zplm}{m}{n}

\usepackage{amsthm}
\usepackage{epsfig}
\usepackage{epsf}
\usepackage{booktabs}
\usepackage{cancel}
\usepackage[normalem]{ulem}
\usepackage{enumerate}

\usepackage{aas_macros}
\usepackage{subfigure}

\def\be{\begin{equation}}
\def\ee{\end{equation}}
\def\bi{\begin{itemize}}
\def\ei{\end{itemize}}
\def\ben{\begin{enumerate}}
\def\een{\end{enumerate}}

\begin{document}
\title{A stochastic search for intermittent gravitational-wave backgrounds}
\author{Jessica Lawrence}
\email{jessica.lawrence@ttu.edu}
\affiliation{Department of Physics, Texas Tech University, Lubbock, TX 79409}
\author{Kevin Turbang}
\email{kevin.turbang@vub.be}
\affiliation{Theoretische Natuurkunde, Vrije Universiteit Brussel, Pleinlaan 2, B-1050 Brussels, Belgium}
\affiliation{Universiteit Antwerpen, Prinsstraat 13, B-2000 Antwerpen, Belgium}
\author{Andrew Matas}
\affiliation{Max Planck Institute for Gravitational Physics (Albert Einstein Institute), D-14476 Potsdam, Germany}
\author{Arianna I. Renzini}
\email{arenzini@caltech.edu}
\affiliation{Department of Physics, California Institute of Technology, Pasadena, California 91125, USA}
\author{Nick van Remortel}
\email{nick.vanremortel@uantwerpen.be}
\affiliation{Universiteit Antwerpen, Prinsstraat 13, B-2000 Antwerpen, Belgium}
\author{Joseph Romano}
\email{joseph.d.romano@ttu.edu}
\affiliation{Department of Physics, Texas Tech University, Lubbock, TX 79409}
\date{\today}

\begin{abstract}
A likely source of a gravitational-wave background (GWB) in the 
frequency band of the Advanced LIGO, Virgo and KAGRA detectors
is the superposition of signals from 
the population of unresolvable stellar-mass binary-black-hole 
(BBH) mergers throughout the Universe.
Since the duration of a BBH merger in band ($\sim\!1~{\rm s}$) is much 
shorter than the expected separation between neighboring mergers
($\sim\!10^3~{\rm s}$), 
the observed signal will be ``popcorn-like"
or intermittent with duty cycles of order $10^{-3}$. However, the standard cross-correlation search for stochastic GWBs currently performed 
by the LIGO-Virgo-KAGRA collaboration is based on a continuous-Gaussian
signal model, which does not take into account the intermittent nature of the background. The latter is better described by a Gaussian mixture-model, which includes a duty cycle parameter that quantifies the degree of intermittence. Building on an earlier paper by Drasco and Flanagan~\cite{Drasco-Flanagan:2003}, 
we propose a stochastic-signal-based 
search for intermittent GWBs. For such signals, this search performs better than the standard continuous cross-correlation search \cite{Allen-Romano:1999}.
We present results of our stochastic-signal-based approach for 
intermittent GWBs applied to simulated data for some simple models, and compare its performance to the other
search methods, both in terms of detection and signal characterization.
Additional testing on more realistic simulated data sets,
e.g., consisting of astrophysically-motivated BBH merger signals 
injected into colored 
detector noise containing noise transients, will be needed 
before this method can be applied with confidence on real gravitational-wave data.
\end{abstract}
\maketitle

\section{Introduction}
\label{s:intro}
The Advanced LIGO~\cite{aLIGO_2015}, Virgo~\cite{aVirgo_2015}, and KAGRA \cite{10.1093/ptep/ptaa125} (LVK) detectors have completed 
their third observing run (O3), increasing the number of
confident detections of gravitational-wave (GW) signals to 
90 overall~\cite{GWTC-3}.
The detected signals are primarily associated with 
stellar-mass binary-black-hole (BBH) mergers,
although a handful of binary-neutron-star (BNS) and 
neutron star-black hole (NSBH) coalescences have 
also been observed~\cite{GW170817, GW200105-GW200115}.
All of these signals are relatively large signal-to-noise
ratio (SNR) events, which stand out above the detector noise
when searched for using matched-filtering techniques~\cite{Helstrom:1968,
Wainstein-Zubakov:1971}.

In addition to these loud, individually-resolvable events, 
the LVK detectors are also being 
showered by GW signals produced by much weaker (e.g., more
distant and/or less massive) sources, 
whose combined effect gives rise to a 
low-level background of gravitational 
radiation---a so-called gravitational-wave background 
(GWB) (see e.g., \cite{Christensen:2018, van_Remortel_2023} and references
cited within).
This background signal is expected to be stochastic
(i.e., random) in the sense that there is no single
deterministic waveform that we can use to perform 
a matched-filter search for this type of GW signal.
Nonetheless, because this signal is present in all
detectors, we can cross-correlate the data from 
multiple detectors to observe the GWB, 
despite its weakness relative to the noise~\cite{Michelson:1987,Allen-Romano:1999}.
Although to date there has not been a direct detection
of a GWB using a stochastic pipeline, we know from Advanced LIGO's and Virgo's 
detections of individual resolvable sources that a background arising from compact binary mergers must exist.
Assuming our detectors are upgraded as planned in the coming years~\cite{2020LRR....23....3A}, and given current projections for the signal~\cite{https://doi.org/10.48550/arxiv.2111.03634}, detecting the GWB may just be a matter of time. On the other hand, we can improve our detection methods to measure this signal sooner. We assume the latter strategy in this paper.

\subsection{Motivation}

A likely source of a GWB in the frequency band of the LVK detectors is the population of stellar-mass BBH mergers 
throughout the Universe. 
Rate estimates
calculated from the BBH signals detected to date~\cite{BBH-properties-O1O2,https://doi.org/10.48550/arxiv.2111.03634}
predict a BBH merger in the observable universe every $\sim 5$-10 minutes on average. 
Since the duration of a BBH merger in the LVK band is of 
order 1~s, the duty cycle $\xi$ of such events 
(defined as the time in band for one merger signal divided by 
the average time between successive mergers) is of order $10^{-3}$. 
Thus, the expected GWB signal is ``popcorn-like” or {\em intermittent}, 
with the signal being ``on” a small fraction of the total 
observation time.
A similar calculation for the population of BNS mergers predicts
(on average) roughly one event every 15 s, while the duration of 
a BNS signal in band is approximately 100~s. 
Thus, BNS merger signals overlap in time leading to a continuous 
(and possibly confusion-limited) background.

The total expected BBH signal is potentially detectable with the Advanced LIGO 
and Virgo detectors when observing at design sensitivity \cite{https://doi.org/10.48550/arxiv.2111.03634,StochImplications:2018}. 
Although the SNRs for the individual events are small, 
the combined SNR of the correlated data summed over all 
events grows like the square-root of the observation time, reaching a 
detectable level of $3\sigma$ (corresponding to a false alarm
probability of approximately $10^{-3}$) after $\sim\!40$ 
months of observation \cite{StochImplications:2018}. 
This estimate of the time-to-detection is based on the standard 
cross-correlation search~\cite{Allen-Romano:1999}, which looks 
for evidence of excess cross-correlated signal power, assuming 
that the amplitude of the GW signal component is 
drawn from a continuous-Gaussian distribution.
This search assumes that the signal is ``on" all the time,
in conflict with the intermittent nature of the stellar-mass
BBH background, which is expected to be the dominant signal.
Thus, although the standard cross-correlation search is 
able to detect the time-averaged signal from an intermittent GWB \cite{Meacher-et-al:2015}, 
this search is sub-optimal
in the sense that the time-to-detection will be longer than that
for a search which properly takes into account the intermittent
nature of the background.

\subsection{Purpose and outline}

The purpose of this paper is to introduce a new 
stochastic-signal-based search that 
specifically targets intermittent GWBs, and 
hence can potentially 
reduce the time-to-detection of the BBH background signal.
This new search is built on the seminal work of Drasco and 
Flanagan~\cite{Drasco-Flanagan:2003}, who proposed a Gaussian mixture-model (GMM) likelihood function for analyzing intermittent GWBs (Sec.~\ref{s:DF}).
Our proposed search for intermittent GWBs looks for excess cross-correlated power in short 
stretches of data. Conversely, a  deterministic-signal-based 
search for the intermittent BBH background was proposed 
by Smith and Thrane~\cite{Smith-Thrane:2018}, which involves marginalizing
over the signal parameters for deterministic BBH chirp waveforms in short ($\sim 4$~s) stretches of data (Sec.~\ref{s:DSI}). 
By construction, our proposed search is adaptable to a generic intermittent GWB since it looks only for excess cross-correlated power.
We also expect our proposed search to be computationally more efficient in detecting a signal than the 
deterministic-signal-based approach of Smith and Thrane, since 
our search ignores the deterministic form of the GW signal waveforms and hence the need to marginalize over all the associated signal parameters.

A brief outline of this paper is as follows: first, we give an overview of the current searches for intermittent GWBs in Sec.~\ref{s:searches}. We then proceed by introducing our proposed stochastic search for intermittent GWBs in Sec.~\ref{s:SSI}. To compare the performance of the various search methods mentioned above, 
we analyze a series of datasets which are tailored to highlight the merits and shortcomings of each style of search. We start in Sec.~\ref{s:Previous results} by considering stationary-Gaussian white 
noise in two co-located and co-aligned detectors, and inject an
intermittent GWB made up of white GW bursts with Gaussian signal amplitudes scaled by distances to the sources drawn from a uniform-in-volume distribution. We then consider a background made up of colored GW bursts\footnote{The term ``burst" will be used throughout this paper as it is the most general, irrespective of the type of signal. In the context of compact binary mergers, these bursts of GWs are often referred to as ``transients".} 
in Sec.~\ref{s:Stochastic bursts}, which follow the expected spectral shape of BBH mergers.
Finally, we analyze a set of
deterministic BBH chirp waveforms in Sec.~\ref{s:Deterministic bursts}, where the chirp parameters are fixed except for the distance to the source, which is also
drawn from a uniform-in-volume distribution.
We conclude in Sec.~\ref{s:discussion} by discussing possible 
extensions of our method and additional tests 
that are needed on more realistic simulated data before it
can be run on real LVK data.

\section{Proposed searches for intermittent GWBs - overview}
\label{s:searches}
The standard continuous cross-correlation search \cite{Allen-Romano:1999} aims to measure the fractional energy density of a GWB, defined as
\begin{equation}
\label{eq:OmegaGW}
    \Omega_{\rm gw} (f) = \frac{1}{\rho_c}\frac{{\rm d}\rho_{\rm gw}}{{\rm d}\ln f},
\end{equation}
where the critical energy density of the Universe is $\rho_c=3H_0^2c^2/(8\pi G)$, $H_0$ is the Hubble constant, $c$ is the speed of light, and
$G$ is Newton’s constant. Alternatively, a GWB can  be characterized by its power spectral density (PSD) $P_{\rm gw}(f)$, which is related to $\Omega_{\rm gw}(f)$ by \cite{Allen-Romano:1999}:
\begin{equation}
\label{eq:OmegaToPSD}
    \Omega_{\rm gw}(f) = \frac{10\pi^2}{3H_0^2}f^3P_{\rm gw}(f).
\end{equation}

For the target signal of a BBH GWB, it is well known that the fractional energy density spectrum is $\Omega_{\rm gw}(f)\propto f^{2/3}$ to good approximation~\cite{TaniaRegimbau_2011}, in the frequency ranges probed by the LVK interferometers. This knowledge can be incorporated into the search, reducing it to the measurement of a single quantity $\Omega_{\rm gw} (f_{\rm ref})$, where $f_{\rm ref}$ is a reference frequency chosen where the sensitivity of the LVK detectors is best (typically 25 Hz)~\cite{O3-isotropic}. For the remainder of the paper, we will refer to $\Omega_{\rm gw} (f_{\rm ref})$ simply as $\Omega_{\rm gw}$ for brevity. For a set of data containing enough events to be statistically significant, $\Omega_{\rm gw}$ is the amplitude of the time and population-averaged energy density. We will refer to this stochastic search for continuous backgrounds described above as SSC.

Since this search assumes a continuous-in-time signal in the data, it does not properly model an important feature of the BBH GWB signal---the intermittency. To remedy this improper modeling, several searches targeting intermittent GWBs specifically have been proposed. We start by giving a high-level overview of these different analysis methods. We refrain from giving details about the actual form of the likelihoods and refer to Appendix~\ref{app:Likelihoods} for more information.

\subsection{Gaussian mixture-model likelihood function for intermittent GWBs}
\label{s:DF}
In 2003, Drasco and Flanagan~\cite{Drasco-Flanagan:2003}
proposed a search for an intermittent GWB that  makes use 
of a GMM likelihood function of the form
\begin{equation}
\label{eq:DFmixture}
{\cal L}_{\rm tot}
=\prod_I^{N_{\rm seg}}
\left[ \xi {\cal L}_{s,I}
+ (1-\xi) {\cal L}_{n,I}
\right]\,,
\end{equation}
where $\xi$ is the probability that a particular segment contains a GW signal,
and ${\cal L}_{s,I}$ and ${\cal L}_{n,I}$ are the likelihood
functions for segment $I$ in the presence and absence of a GW signal, i.e., the signal and noise likelihoods.
For the simple toy model considered in their paper (i.e., 
single-sample GW ``bursts", occurring with probability $\xi$ drawn from a fixed Gaussian distribution 
with variance $\sigma_b^2$, and injected into uncorrelated white 
noise in two co-located and 
co-aligned detectors),
the signal and noise parameters that enter the likelihood functions 
${\cal L}_{s,I}$ and ${\cal L}_{n,I}$ 
are the variances 
$(\sigma_b^2, \sigma^2_{n_1}, \sigma^2_{n_2})$
and $(\sigma^2_{n_1}, \sigma^2_{n_2})$,
respectively.
Single-sample bursts are bursts whose duration is less than the sample period $\Delta t$.
By maximizing ${\cal L}_{\rm tot}$ with respect to all four 
parameters $(\xi, \sigma^2_b, \sigma^2_{n_1}, \sigma^2_{n_2})$, 
Drasco and Flanagan obtained a detection statistic
(the maximum-likelihood statistic), 
which they could use to search for intermittent GWBs. Note that in the case $\xi=1$, i.e., assuming the signal is always present, one recovers the standard continuous-Gaussian search introduced above.

Although Drasco and Flanagan tested their proposed method with a test statistic within a frequentist framework, we have decided to work within a Bayesian framework in this paper. We define several concepts of importance within this framework before moving on to the discussion of the results of Drasco and Flanagan. 

Given a likelihood function ${\cal L}_{\rm tot}$ and priors $\pi$, the joint posterior distribution for the duty cycle 
and the signal+noise parameters can be computed using Bayes' theorem:
\be
p(\xi,\sigma_b^2, \sigma^2_{n_1}, \sigma^2_{n_2}|d)
=\frac{{\cal L}_{\rm tot}(d|\xi,\sigma_b^2, \sigma^2_{n_1}, \sigma^2_{n_2})
\pi(\xi)\pi(\sigma_b^2)\pi(\sigma^2_{n_1})\pi( \sigma^2_{n_2})}
{{\pazocal Z}(d)}\,,
\ee
where 
\be
{\pazocal Z}(d)
\equiv 
\int{\rm d}\xi\int{\rm d}\sigma_b^2\int{\rm d}\sigma_{n_1}^2\int{\rm d}\sigma_{n_2}^2\>
{\cal L}_{\rm tot}(d|\xi,\sigma_b^2, \sigma^2_{n_1}, \sigma^2_{n_2})
\pi(\xi)\pi(\sigma_b^2)\pi(\sigma^2_{n_1})\pi( \sigma^2_{n_2})
\ee
is the model evidence.
Marginalized posterior distributions (for each parameter
separately) are obtained by integrating the joint posterior 
distribution over all the other parameters, e.g.,
\be
p(\xi)
=\int{\rm d}\sigma_b^2\int{\rm d}\sigma_{n_1}^2\int{\rm d}\sigma_{n_2}^2\>
p(\xi,\sigma_b^2, \sigma^2_{n_1}, \sigma^2_{n_2})\,.
\ee
Of course, likelihood functions, priors, etc., are all
calculated in the context of a particular
choice of analysis model ${\cal M}_\alpha$ 
(e.g., a GMM likelihood
search for intermittent GWBs or the standard continuous-Gaussian search), 
which we have not indicated in the above expressions.
If we explicitly denote the dependence of the above 
distributions 
on the choice of analysis model, we can define the 
Bayes factor between models ${\cal M}_\alpha$ 
and ${\cal M}_\beta$ as
\be
\pazocal{B}_{\alpha\beta}(d) \equiv 
\frac{{\pazocal Z}(d|{\cal M}_\alpha)}
{{\pazocal Z}(d|{\cal M}_\beta)}\,.
\label{e:bayes_factor}
\ee
Assuming equal prior odds for the two models, the Bayes factor
tells us how much more the data favors model ${\cal M}_\alpha$ 
relative to ${\cal M}_\beta$.
Throughout this paper, we will make plots of the natural logarithm of the 
Bayes factor as a function of the duty cycle to compare the
various search methods.

With these concepts in mind, we now move to the discussion of the results of the proposed GMM likelihood. Drasco and Flanagan showed that their detection statistic for intermittent 
GWBs performs better than the standard cross-correlation statistic for continuous-Gaussian backgrounds when the duty cycle $\xi$ is sufficiently small. To illustrate this, we implement their proposed GMM likelihood in a Bayesian framework. Instead of using their proposed frequentist detection statistic, we use the Bayes factor as a measure of efficiency. To be able to study its behavior as a function of the duty cycle, we combine 100 data realizations for each $\xi$ value. Each data realization consists of 40,000 segments, where a fraction of them contains single-sample bursts drawn from a Gaussian distribution with variance $\sigma^2_b=1$. 

\begin{figure}[hbtp!]
\begin{center}
\includegraphics[clip=true, angle=0, width=0.5\textwidth]{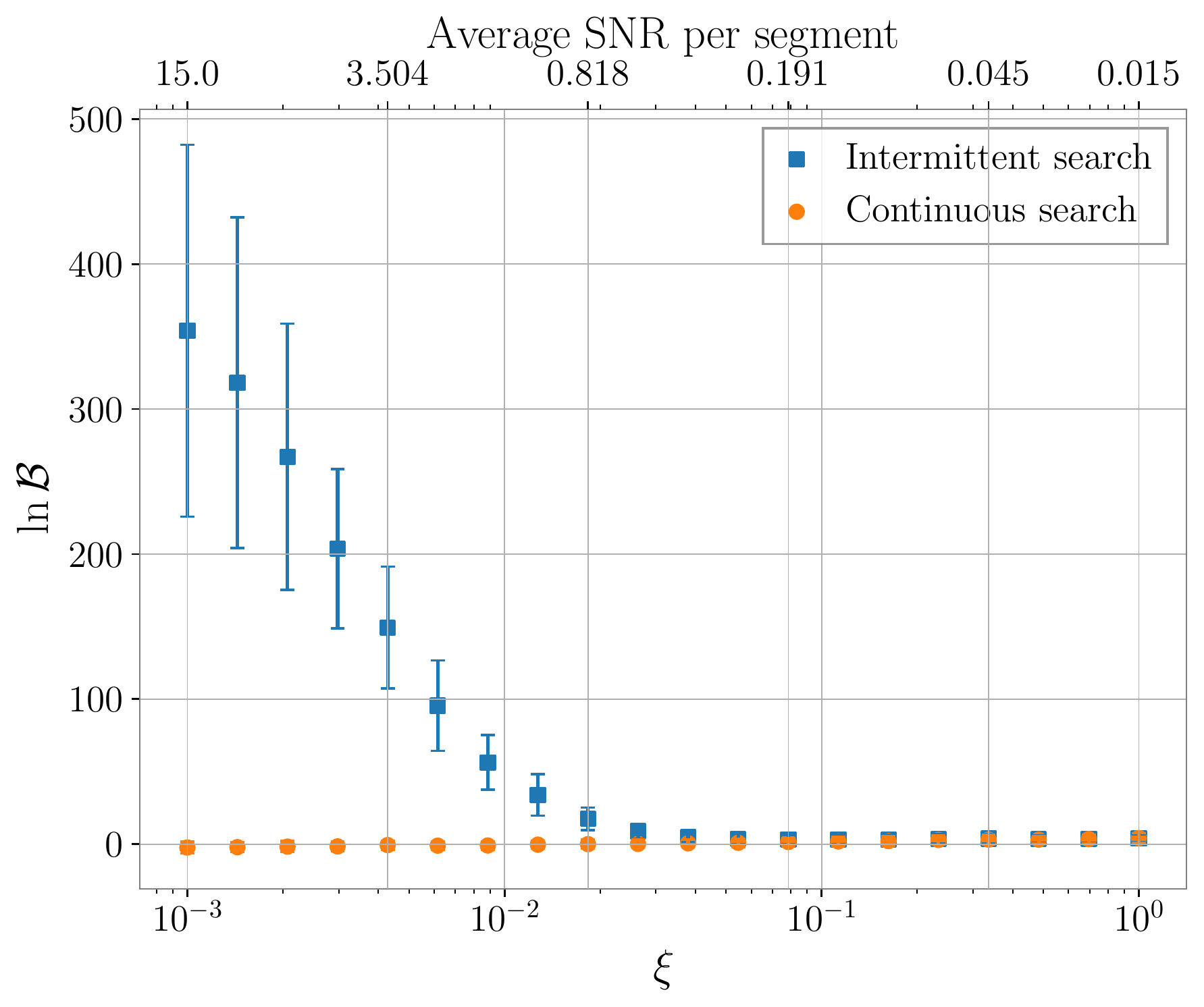}
\caption{ln Bayes factors of the signal+noise model to the noise-only model as a function of the duty cycle $\xi$ for the intermittent search (blue) and the continuous search (orange) where the signal consists of single sample bursts drawn from a Gaussian distribution of variance $\sigma_b^2$.}
\label{f:lnBFDF}
\end{center}
\end{figure}

We keep the total continuous-Gaussian signal-to-noise ratio fixed to 3, computed using \eqref{e:rho_DF} and \eqref{e:rho_seg}, by adjusting the noise variances for each value of the duty cycle, rather than adjusting the signal parameters. So, as $\xi$ decreases, the segment signal-to-noise ratios must increase, which means that the noise variances must decrease. This is illustrated in Fig. \ref{f:lnBFDF}, where both the continuous-in-time, i.e., $\xi=1$ in \eqref{eq:DFmixture}, and the intermittent GMM likelihood analysis methods are used. Each plotted point corresponds to the mean of the ln Bayes factor over 100 realizations of data, while the error bars correspond to the standard deviation of the ln Bayes factor.

While the continuous search performs equally well for all duty cycles (since it assumes $\xi=1$), the Bayes factor 
for the GMM likelihood
increases as $\xi$ decreases, exceeding the 
continuous stochastic likelihood Bayes factor, illustrating that the GMM likelihood performs better than the continuous likelihood for smaller values of $\xi$.
Equivalently, the relative performance of the Bayes factors 
shown in Fig.~\ref{f:lnBFDF} can be expressed in terms of
\be
\label{e:rho_DF}
\rho_{\rm seg}\equiv \frac{\sigma_b^2}{\sigma_{n_1}\sigma_{n_2}}\,,
\ee
which is the expected signal-to-noise ratio in an individual segment assuming the presence of a GW signal with burst variance $\sigma^2_b$.
In terms of $\rho_{\rm seg}$, the condition for the GMM likelihood to perform better than the continuous likelihood is
\begin{equation}
\rho_{\rm seg} \sim 1\,.
\end{equation}
In the limit where $\rho_{\rm seg}\ll 1$, 
the GW signals in an individual
segment are sufficiently weak that the GMM likelihood does not perform any better than the standard stochastic
continuous likelihood.
Conversely, when $\rho_{\rm seg}\gg 1$, the GW signals in the individual segments 
are so strong that they are individually resolvable, with
segment signal-to-noise ratios exceeding the threshold needed for 
detection with a single-detector burst statistic.
In other words, a search for an intermittent GWB is the most 
sensitive search when 
the GW signals in the individual segments are marginally sub-threshold
($\rho_{\rm seg}\sim 1$).

Furthermore, we can determine an approximate value of 
$\rho_{\rm seg}$ for the LVK detectors,
for the population of stellar-mass BBH mergers throughout 
the Universe.
As mentioned in Sec.~\ref{s:intro}, it should take $\sim 40$ months of observation using the 
standard continuous-Gaussian cross-correlation statistic to 
observe the BBH background 
with a total signal-to-noise ratio $\rho_{\rm tot}=3$ \cite{GW170817-stoch-implications:2018}.
Since the segment duration proposed by Smith and 
Thrane~\cite{Smith-Thrane:2018} 
for an intermittent search is of order $T_{\rm seg}\sim 4~{\rm s}$
(see Sec.~\ref{s:DSI} for more details), 40~months of 
observation corresponds to $N_{\rm seg}\sim 2.5\times 10^7$~segments.
The final input that we need to  do the calculation is the 
expected duty cycle of the signal, which for stellar-mass 
BBH mergers throughout the Universe is $\xi\sim 10^{-3}$.
These values imply
\be
\rho_{\rm seg} =\frac{\rho_{\rm tot}}{\xi\sqrt{N_{\rm seg}}}\sim 0.6\,,
\label{e:rho_seg}
\ee
which is in the regime where a search for an intermittent GWB should 
start to perform better than the standard continuous-Gaussian 
cross-correlation search. The value of $\rho_{\rm seg}$ at which the intermittent search begins to outperform the continuous search in Fig.~\ref{f:lnBFDF} matches this result.


\subsection{Deterministic-signal-based search for intermittent GWBs}
\label{s:DSI}

In 2018, Smith and Thrane~\cite{Smith-Thrane:2018}
extended the work of Drasco and Flanagan~\cite{Drasco-Flanagan:2003} 
by proposing an optimal fully-Bayesian deterministic-signal-based
search for the intermittent GWB produced by the population 
of stellar-mass BBH mergers throughout the Universe.
As in \cite{Drasco-Flanagan:2003}, Smith and Thrane~\cite{Smith-Thrane:2018}
assume a mixture model for the intermittent GW signals. They chose a segment duration $\sim\!4$~s, which is long enough to 
include a typical BBH chirp signal, yet short enough that the probability 
of two such signals occurring in a single segment is negligibly small
($\sim 10^{-4}$).
However, instead 
of considering single-sample GW bursts drawn from a fixed 
Gaussian distribution, they considered finite-duration  deterministic BBH chirp waveforms
$h=h_{\rm chirp}(t;\theta)$, where $\theta$ are the chirp 
parameters (e.g., the component masses and spins of the two
BHs, the inclination angle of the orbital plane relative
to the line of sight, etc). 
Smith and Thrane then marginalized (instead of maximized) over the 
signal parameters for each segment of data, assuming prior probability 
distributions for these parameters, while replacing the noise 
parameters by measured estimates of these quantities.
If the signal priors are conditioned on 
segment-independent population parameters $\theta_{\rm pop}$, 
which parameterize the distributions from which
the individual masses, spins, etc., are drawn,
then the final (marginalized) likelihood function 
${\cal L}_{\rm tot}\equiv{\cal L}_{\rm tot}(d|\xi,\theta_{\rm pop})$
depends only on 
the duty cycle $\xi$ and the population parameters $\theta_{\rm pop}$.
Finally, doing Bayesian inference calculations given 
${\cal L}_{\rm tot}$ and a prior for $\xi$ and $\theta_{\rm pop}$, 
Smith and Thrane were able to construct joint posterior 
distributions for $\xi$ and $\theta_{\rm pop}$ as well as Bayes 
factors comparing the evidence for this intermittent signal model 
and e.g., that for the standard cross-correlation search for a 
continuous-Gaussian GWB.

The deterministic-signal-based search of Smith and Thrane
is expected to decrease the time-to-detection of the intermittent
GWB produced by stellar-mass BBH mergers by a factor 
of $\sim\!1000$ relative to the 
standard continuous-Gaussian search~\cite{Smith-Thrane:2018},
by taking into account both the intermittent nature of the 
signal as well as knowledge of the form of the individual 
waveforms, whose parameters are marginalized over.
For this factor of $\sim 1000$ determination, 
they did not consider any
population parameters, so the only parameter that they
needed to infer from the data was the duty cycle $\xi$.
A posterior distribution for $\xi$ sufficiently bounded
away from zero would be evidence of a confident detection of 
an intermittent GWB signal.
The gain in time-to-detection comes at the computational
cost of having to perform Bayesian marginalization over all the
BBH chirp signal parameters for every 4~s segment of data.
This search is currently in the testing phase, in 
preparation for running on real LVK 
data in the near future.

Within this paper, for comparative purposes, we will implement a much simpler version of this deterministic-signal-based search. We will use the acronym DSI throughout this work to refer to the deterministic-signal-based search for intermittent GWBs.

\section{SSI: Stochastic search for intermittent GWBs}
\label{s:SSI}
Building off the work of Drasco and Flanagan~\cite{Drasco-Flanagan:2003},
we propose a new search 
based on a stochastic-signal model consisting 
of intermittent ``bursts" of correlated stochastic GWs with 
unknown duty cycle $\xi$, in otherwise uncorrelated noise
in two detectors. We call this search SSI, for \underline{s}tochastic \underline{s}earch for \underline{i}ntermittent GWBs, referencing both the signal model the analysis assumes, as well as the type of background for which it is designed.
To make the connection with BBH mergers, we assume 
that these bursts of GWs last on the order of a few seconds 
so the data are split into short stretches as in
Smith and Thrane, and that the power spectrum
in the LVK detectors 
goes like $f^{-7/3}$, appropriate for binary inspiral. This corresponds to a fractional energy density
spectrum $\Omega_{\rm gw}(f)\propto f^{2/3}$, as introduced in \eqref{eq:OmegaGW}.

Rather than marginalize over the parameters of 
deterministic BBH chirp waveforms as in the 
deterministic-signal-based approach, our search looks for 
excess cross-correlated power when the signal 
is assumed to be present, using a mixture-model likelihood
function.
Thus, we trade off optimality for computational 
efficiency and flexibility relative to the 
deterministic-signal-based approach, while still 
accounting for the intermittent nature of
the BBH background, which is missing from the standard
cross-correlation search for continuous-Gaussian GWBs.

We begin by dividing up the data into short segments such that the probability of a segment containing more than one signal is small.
The total likelihood is given by a product over segments of the GMM likelihood function
\be 
{\cal L}_{\rm tot}(d|\xi, \theta_{s,{\rm pop}}, \theta_{n}) = \prod_I
\left[\xi {\cal L}_s(d_I|\theta_{s,{\rm pop}},\theta_n)
+ (1-\xi){\cal L}_n(d_I|\theta_n)\right],
\label{e:SSI_mixture}
\ee
where $\theta_n$ represents the noise parameters, $\theta_{s,{\rm pop}}$ represents the signal population parameters, and $d_I$ represents the data in segment $I$.

For our stochastic-signal-based search, the segment-dependent signal likelihood takes the form
\be 
{\cal L}_s(d_I|\theta_{s,{\rm pop}}, \theta_n)
\equiv \int{\rm d}\theta_{s,I}\>
{\cal L}_n(d_I|\theta_{s,I},\theta_n)
\pi(\theta_{s,I}|\theta_{s,{\rm pop}})\,,
\label{e:stochastic}
\ee
where the segment-dependent signal parameters $\theta_{s,I}$ are marginalized over. Marginalizing over the correct segment prior is an important and necessary step in order to recover correct and unbiased results.

We choose to write the likelihood for a specific set of parameters, $\theta_{s,{\rm pop}}=\langle\Omega_b\rangle$, $\theta_{s,I}=\Omega_{b,I}$, and $\theta_{n}=\{\sigma^2_{n_1}, \sigma^2_{n_2}\}$, where $\langle\Omega_b\rangle$ is the population-averaged energy density amplitudes of bursts of GW power and $\Omega_{b,I}$ is the energy density amplitude in data segment $I$. The population parameter $\langle\Omega_b\rangle$ is related to $\Omega_{\rm gw}$, introduced at the beginning of Sec. \ref{s:searches}, by:
\be
\Omega_{\rm gw} = \xi \langle\Omega_b\rangle.
\label{e:Omega_gw}
\ee
Recall that $\Omega_{\rm gw}$ is what the standard cross-correlation search for a continuous-Gaussian GWB estimates.
For the analyses included in this paper, we simulate stationary, white-Gaussian noise. This means that the power spectrum of the noise is independent of frequency and has the value 
\be
P_{n_\mu}=\frac{\sigma_{n_\mu}^2}{f_{\rm high}-f_{\rm low}}
\label{e:noise_power}
\ee
where $\mu=1,2$ is the detector index and $f_{\rm low}$ and $f_{\rm high}$ are the low- and high-frequency
cutoffs for our search.
We will take $f_{\rm high}$ to equal the Nyquist 
critical frequency $f_{\rm nyq} \equiv 1/(2\Delta t)$, 
where $\Delta t$ is the sample period. Each segment of time-domain data of duration $T$ is Fourier transformed and coarse-grained to frequencies $f_k$ having frequency resolution $M/T$.
We then take our noise parameters to be the variance of the noise in each detector. Under these assumptions, the segment-dependent signal likelihood~\eqref{e:stochastic} becomes
\begin{align}
{\cal L}_s(d_I|\langle\Omega_{b}\rangle,\sigma^2_{n_1}, \sigma^2_{n_2}) &= 
\int d\Omega_{b,I} \pi(\Omega_{b,I}|\langle\Omega_{b}\rangle)\prod_k
\frac{1}{(\pi T/2)^{2M}(P_{1,I}(f_k)P_{2,I}(f_k) - P^2_{b,I}(f_k))^{M}}
\nonumber\\
& \times
\exp\left\{-\frac{M}
{(P_{1,I}(f_k)P_{2,I}(f_k) - P^2_{b,I}(f_k))}\left[
\hat P_{1,Ik}P_{2,I}(f_k) 
+ \hat P_{2,Ik}P_{1,I}(f_k)
- 2 \hat P_{b,Ik} P_{b,I}(f_k)
\right]\right\}\,,
\label{e:likefinal_colored_nonstationary-signal}
\end{align}
where
\be
P_{1,I}(f)
\equiv \frac{\sigma_{n_1}^2}{f_{\rm high}-f_{\rm low}} + P_{b,I}(f)\,,
\qquad
P_{2,I}(f)
\equiv \frac{\sigma_{n_2}^2}{f_{\rm high}-f_{\rm low}} + P_{b,I}(f)\,,
\qquad
P_{b,I}(f)
\equiv \Omega_{b,I} H(f),
\ee
are the total auto-correlated power spectra in each detector and the power spectrum for a GW burst in segment $I$, and $k$ runs over the coarse-grained frequencies $f_k$. The spectral shape $H(f)$ is of the form
\be
H(f)\equiv \frac{3H_0^2}{10\pi^2}\frac{1}{f_{\rm ref}^3}
\left(\frac{f}{f_{\rm ref}}\right)^{-7/3}\,.
\label{e:H(f)}
\ee
The Fourier transformed data enter the evidence via the following quadratic 
combinations
\be
\begin{aligned}
&\hat P_{1,Ik}
\equiv \frac{2}{T}\,\frac{1}{M}\sum_{k'=k-M/2T}^{k+M/2T-1}
|\tilde d_{1,Ik'}|^2\,,
\\
&\hat P_{2,Ik}
\equiv \frac{2}{T}\,\frac{1}{M}\sum_{k'=k-M/2T}^{k+M/2T-1}
|\tilde d_{2,Ik'}|^2\,,
\\
&\hat P_{b,Ik}
\equiv \frac{2}{T}\,\frac{1}{M}\sum_{k'=k-M/2T}^{k+M/2T-1}
{\rm Re} \left(\tilde d_{1,Ik'}^* \tilde d_{2,Ik'}^{}\right)\,,
\label{e:SSI-coarse-grained-estimators}
\end{aligned}
\ee
which are coarse-grained estimators (i.e., averaged over fine-grained frequencies labeled by $k'$) of the total auto-correlated and
cross-correlated power spectra in the two detectors.

The segment-dependent noise likelihood can similarly be written as 
\be
{\cal L}_n(d_I|\sigma^2_{n_1},\sigma^2_{n_2})
=\prod_k
\frac{1}{(\pi T/2)^{2M}\left(P_{n_1}(f_k)P_{n_2}(f_k)\right)^M}
\exp\left\{-M
\left[
\frac{\hat P_{1,Ik}}{P_{n_1}} 
+\frac{\hat P_{2,Ik}}{P_{n_2}}
\right]\right\}\,.
\label{e:likefinal_colored_nonstationary-noise}
\ee

In principle, the noise parameters 
$\theta_n=\{\sigma_{n_1}^2,\sigma_{n_2}^2\}$ in the likelihood functions above  should be inferred 
together with the signal population parameters $\theta_{s,{\rm pop}}=\langle\Omega_b\rangle$,
as part of the Bayesian inference procedure. Doing so defines the so-called {\em full} version of the analyses.
However, as LVK noise is stationary to good approximation, it is typically sufficient to use measured estimates of the noise parameters (denoted by $\bar\sigma_{n_1}^2$ and $\bar\sigma_{n_2}^2$ and computed using \eqref{e:noisepower-reduced}) in 
the likelihood function, thereby avoiding having to infer them in this analysis. 
We refer to this approach as the {\em reduced} form of the analyses, which is 
computationally cheaper than the full form. The reduced version of the likelihood requires that the 
cross-correlation estimators be approximately Gaussian,
which holds only if the number of samples per segment
$N$ is sufficiently large. 

The reduced segment-dependent signal likelihood is given by \cite{Matas-Romano:2021}:
\be
{\cal L}_s(d_I|\langle\Omega_{b}\rangle,\bar{\sigma}^2_{n_1}, \bar{\sigma}^2_{n_2}) = \int d\Omega_{b,I} \pi(\Omega_{b,I}|\langle\Omega_{b}\rangle) \frac{1}{\sqrt{2\pi\,{\rm var}(\bar\Omega_{b,I})}} 
\exp\left[-\frac{(\hat\Omega_{b,I}-\Omega_{b,I})^2}{2\,{\rm var}(\bar\Omega_{b,I})}\right]\,,
\label{e:SSI-reduced-colored-Zs}
\ee
where
\be
\hat \Omega_{b,I}
\equiv 
\frac{\sum_k Q_I(f_k)\hat  P_{b,Ik}}
{\sum_{k'} Q_I(f_{k'}) H(f_{k'})}\,,
\qquad
{\rm var}(\bar\Omega_{b,I})
\equiv \left({2M}\sum_k 
Q_I(f_k)H(f_k)\right)^{-1}
\label{e:optimally-filtered-CC}
\ee
are the optimally-filtered cross-correlation estimators and
corresponding variances, which are constructed from 
coarse-grained estimates of the cross-correlated power $\hat P_{b,Ik}$ (given by \eqref{e:SSI-coarse-grained-estimators})
and the segment-dependent optimal filter function
\be
Q_I(f)
\equiv \frac{H(f)}{\bar P_{1,I}(f)\bar P_{2,I}(f)}\,,
\ee
where
\be
\bar P_{1,I}(f)\equiv 
\frac{\bar{\sigma}_{n_1}^2}{f_{\rm high}-f_{\rm low}}
 +\Omega_{b,I}H(f)\,,
\qquad
\bar P_{2,I}(f)\equiv
\frac{\bar{\sigma}_{n_2}^2}{f_{\rm high}-f_{\rm low}}
 +\Omega_{b,I}H(f)\,.
\ee
Note that $Q_I(f)$ is a generalization of the
standard optimal filter for an $f^{-7/3}$ power spectrum
(see e.g.,~\cite{Allen-Romano:1999, Romano-Cornish:2017}),
extended to include the segment-dependent burst 
contribution, i.e., dependent on the likelihood parameter $\Omega_{b,I}$, to the total auto-correlated power estimates
$\bar P_{1,I}(f)$, $\bar P_{2,I}(f)$.

The reduced segment-dependent noise likelihood 
${\cal L}_n(d_I|\bar{\sigma}^2_{n_1}, \bar{\sigma}^2_{n_2})$ is given by
\begin{align}
{\cal L}_n(d_I|\bar{\sigma}^2_{n_1}, \bar{\sigma}^2_{n_2})
&=
\frac{1}{\sqrt{2\pi\,{\rm var}(\bar\Omega_{b})}} 
\exp\left[-\frac{(\hat\Omega_{b,I})^2}{2\,{\rm var}(\bar\Omega_{b})}\right]\,,
\label{e:SSI-reduced-colored-Zn}
\end{align}
where $\hat\Omega_{b,I}$ and ${\rm var}(\bar\Omega_b)$ are
the same as for the segment-dependent signal likelihood, but
with a segment-{\em independent}, noise-only optimal filter function
\be
Q(f)\equiv \frac{H(f)}{\bar P_{n_1}\bar P_{n_2}}\,.
\ee

\section{Analyses}
\label{s:toymodels}

In this section, we describe in detail a set of analyses, which we use to illustrate
various aspects of the search methods described above.
The tests that these analyses allow us to perform should 
be thought of as providing a 
``proof-of-principle" demonstration of our proposed 
stochastic-signal-based search for intermittent GWBs.
A more rigorous test of this search on actual LVK
noise and realistic injected BBH chirp signals is a 
topic for future investigation 
(see Sec.~\ref{s:discussion} for more details).

For all the analyses we consider, we assume 
white, stationary-Gaussian noise in two co-located and 
co-aligned detectors with variances $\sigma^2_{n_1}$ and
$\sigma^2_{n_2}$, respectively.
The assumption of co-located and co-aligned detectors
means that we can ignore the so-called 
overlap reduction function~\cite{Christensen:1992,Flanagan:1993},
which encodes the reduction in cross-correlated
power that comes from correlating two physically separated
and possibly misaligned detectors.
To calculate the total SNR for each set of data, we use the average SNR per segment computed using formulas specified below for each data set and rearrange \eqref{e:rho_seg} to solve for $\rho_{\rm tot}$. We note that this $\rho_{\rm tot}$ is the total SNR of the continuous-in-time cross-correlation search, which assumes the signal exists in every segment of data. For our intermittent analyses, we use this definition of total SNR to quantify the strength of the GW signal.

\subsection{Extension of previous work}
\label{s:Previous results}

In Section \ref{s:DF}, the results of Drasco and Flanagan~\cite{Drasco-Flanagan:2003} are reproduced within a Bayesian framework (see Fig. \ref{f:lnBFDF}). We remind the reader that the signals considered there are single-sample GW ``bursts" drawn from a fixed Gaussian distribution 
with variance $\sigma_b^2$. We proceed with the generalisation of the proposed GMM likelihood to allow for more realistic signals. 

As a first step, we now allow multi-sample ($N\gg 1$) bursts of white stochastic GWs having duty cycle $\xi$,
with signal samples drawn from
a probability distribution that depends on the 
distance $r$ to an individual source.
For a source at arbitrary
reference distance $r_{\rm ref}$, we draw the signal 
samples from a Gaussian distribution with fixed 
variance $\sigma^2_{\rm ref}$. For a source at a general distance $r$, we 
first draw the signal samples from a Gaussian distribution with
variance $\sigma^2_{\rm ref}$
as explained above, and then rescale the 
samples by a factor of $r_{\rm ref}/r$, since 
GW signal amplitudes fall off as $1/r$ \cite{Maggiore-book}.
Thus,
\be
\sigma^2_b(r) \equiv \sigma^2_{\rm ref}\,\frac{r^2_{\rm ref}}{r^2}
\label{e:sigma2b}
\ee
is the burst variance for a source at distance $r$. 

For the population model, we will assume that the
source distances are drawn from a {\em uniform-in-volume}
probability distribution
\be
p(r|r_{\rm max})
\equiv\frac{3r^2}{r_{\rm max}^3 - r_{\rm min}^3}\,,
\label{e:unif-in-vol}
\ee
where $r_{\rm max}$ is the maximum distance out to which the sources are formed (i.e., an unknown population 
parameter that will eventually be inferred from the data). The parameter $r_{\rm min}$ is taken to be a fixed, known parameter, for simplicity. Note that choosing $r_{\rm min}\neq 0$ in the simulation process limits the number of GW bursts that are so loud that they are individually detectable in a single segment of data. We also note that this choice of population model is a simplification as it does not take into account cosmology.

It follows from \eqref{e:sigma2b} and \eqref{e:unif-in-vol} that
\be
p(\sigma^2_b(r)|r_{\rm max}) = 
\frac{3r_{\rm ref}^3}{2(r_{\rm max}^3- r_{\rm min}^3)}
(\sigma^2_{\rm ref})^{3/2}(\sigma^2_b(r))^{-5/2}
\label{e:p_sigma2b}
\ee
is the probability distribution for the signal
variance $\sigma^2_b(r)$ associated with a source at distance $r$.
We also define the population-averaged burst variance:
\be
\langle\sigma^2_b\rangle
\equiv
\int_{r_{\rm min}}^{r_{\rm max}}{\rm d}r\>
p(r|r_{\rm max})\sigma^2_b(r)
= 3\sigma^2_{\rm ref} 
\frac{r^2_{\rm ref}(r_{\rm max}-r_{\rm min})}{r_{\rm max}^3 - r_{\rm min}^3}\,,
\label{e:sigma2b_popavg}
\ee
which is obtained by averaging $\sigma^2_b(r)$ over the 
uniform-in-volume-distributed source distances $r$.
We define 
$\sigma^2_{\rm gw} \equiv \xi\langle\sigma^2_b\rangle$, 
which has the interpretation 
of being the time and population-averaged variance
of the signals. This quantity is what the 
standard cross-correlation search for a 
continuous-Gaussian GWB (SSC) estimates.

Since the probability distribution for $\sigma^2_b(r)$ depends on
just  one free parameter, i.e., $r_{\rm max}$ in \eqref{e:p_sigma2b}, 
we can equally well
use the population-averaged variance $\langle\sigma^2_b\rangle$ 
as the population parameter for the probability distribution.
Solving \eqref{e:sigma2b_popavg} for $r_{\rm max}$ in terms of
$\langle\sigma^2_b\rangle$, we find
\be
\begin{aligned}
&r_{\rm max} = r_{\rm min}\left(
\sqrt{-\frac{3}{4}+ 3 \frac{\sigma^2_{b,{\rm max}}}{\langle\sigma^2_b\rangle}}
-\frac{1}{2}\right)\,,
\\
&\sigma^2_{b,{\rm max}} \equiv \sigma^2_b(r_{\rm min})
=\sigma^2_{\rm ref}\frac{r^2_{\rm ref}}{r^2_{\rm min}}\,,
\label{e:rmax_sigma2b}
\end{aligned}
\ee
leading to
\be
p(\sigma^2_b(r)|\langle\sigma^2_b\rangle)= 
\frac{\langle\sigma^2_b\rangle (\sigma^2_{b,{\rm max}})^{1/2}}
{\sqrt{-3 + 12{\sigma^2_{b,{\rm max}}}/{\langle\sigma^2_b\rangle}}-3}\,
(\sigma^2_b(r))^{-5/2}\,.
\label{e:p_sigma2b_popavg}
\ee
The above expression is somewhat messy, but it will be useful 
when we perform Bayesian inference on 
$\langle\sigma^2_b\rangle$.
Building on the above, we define the average segment SNR of the distribution in a similar manner as \eqref{e:sigma2b_popavg},
\be
\langle\rho_{\rm seg}\rangle = \int_{r_{\rm min}}^{r_{\rm max}}{\rm d}r\>
p(r|r_{\rm max}) \rho_{\rm seg}(r)
\ee
where $\rho_{\rm seg}(r)$ for these signals is given by \eqref{e:rho_DF} with $\sigma_{b}^2$ replaced  by $\sigma_{b}^2(r)$.

\begin{figure}[hbtp!]
\begin{center}
\includegraphics[clip=true, angle=0, width=0.49\textwidth]{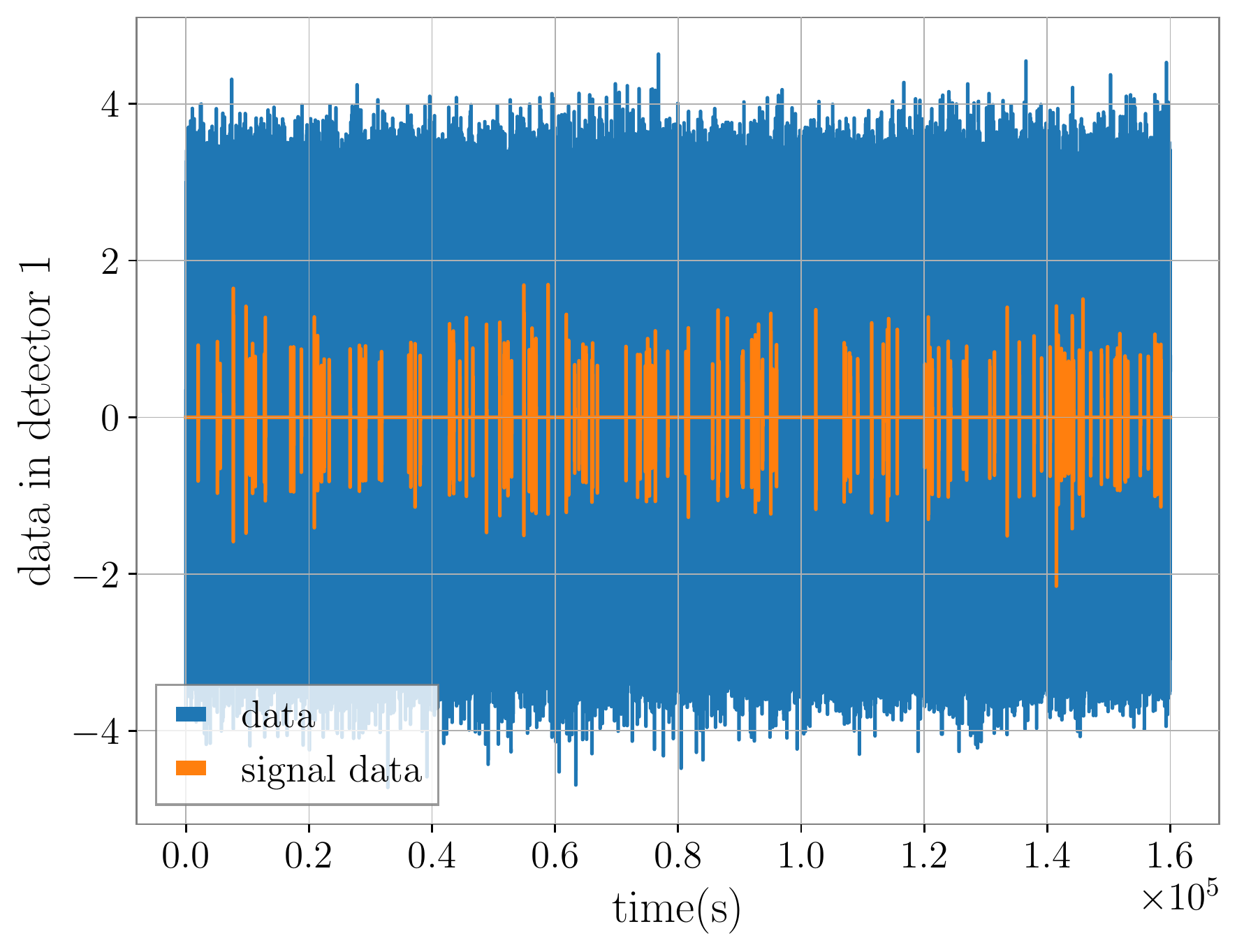}
\includegraphics[clip=true, angle=0, width=0.46\textwidth]{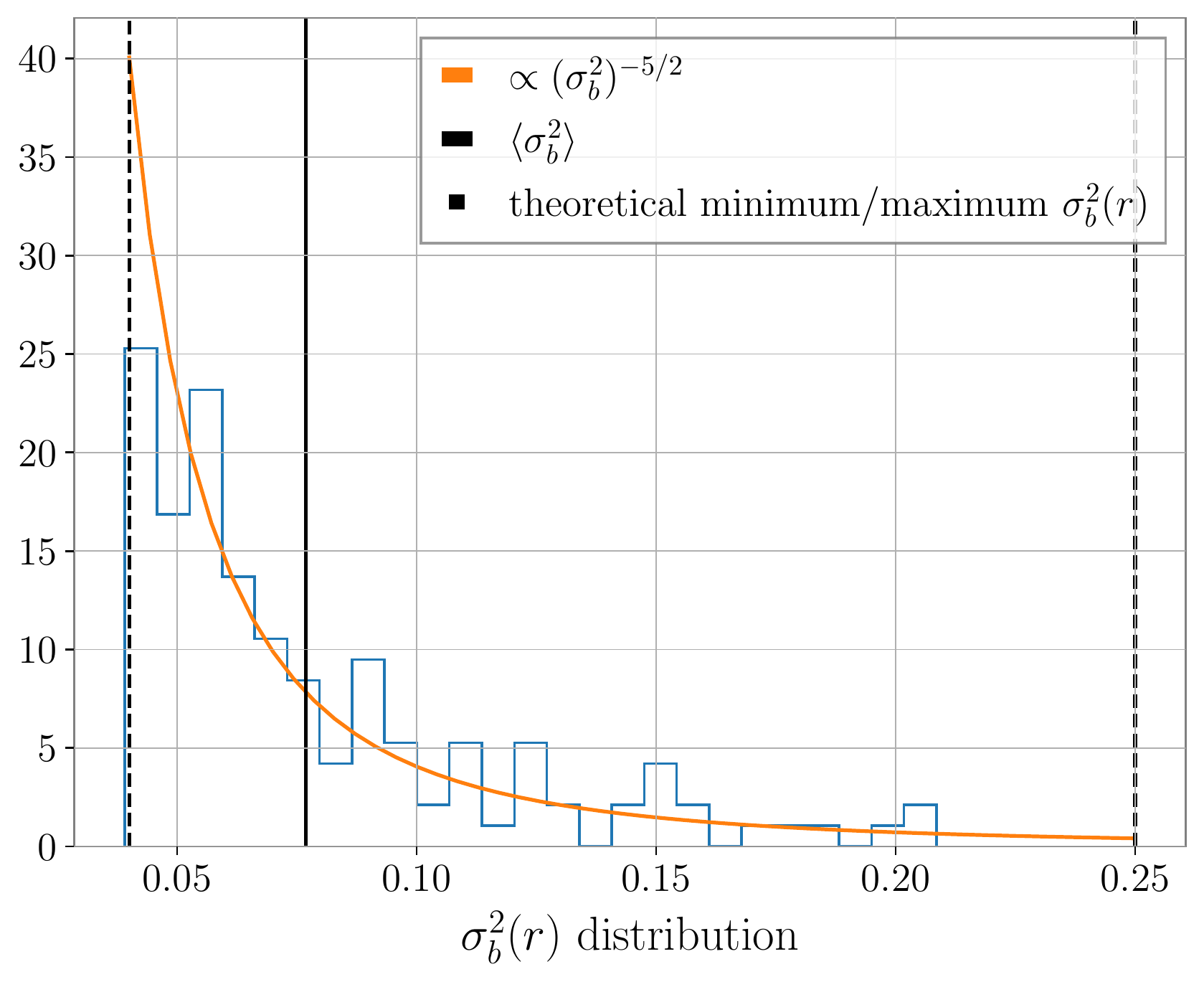}
\caption{\textbf{Left:} Example of simulated data with amplitudes drawn from a uniform-in-volume distribution. The parameters used for this injection are given in the `Extension of previous work' section of Table~\ref{t:parameter table}. \textbf{Right:} Distribution of the burst variances drawn from a uniform-in-volume distribution, with theoretical minimum and maximum burst variance evaluated at $r_{\rm max}$ and $r_{\rm min}$, respectively, and average burst variance $\langle\sigma_b^2\rangle$ computed according to \eqref{e:sigma2b_popavg}.}
\label{f:toymodel_2_data}
\end{center}
\end{figure}

\begin{table}[h!]
    \centering

\begin{tabular}{ |p{1cm}|p{1cm}|p{1.5cm}|p{1cm}|p{1cm}|p{1cm}|p{1cm}|p{1cm}|p{1cm}|p{1cm}|p{1cm}|}
 \hline
 \multicolumn{11}{|c|}{Extension of previous work}
 \\
 \hline
 \hline
 $N_{\rm seg}$ &$N$& $\xi$&$r_{\rm min}$ &$r_{\rm max}$ &$r_{\rm ref}$ &$\sigma^2_{\rm ref}$ &  $\langle\sigma_b^2\rangle$& $\sigma^2_n$ & $\langle\rho_{\rm seg}\rangle$ & $\rho_{\rm tot}$\\
 \hline
 4$\times 10^4$ & 2048 & 2.98$\times 10^{-3}$ & 2 & 5 & 1 & 1 & 0.0769 & 0.691 & 5.04 & 3\\
 \hline
\end{tabular}
\\~\\~\\
\begin{tabular}{ |p{1cm}|p{1cm}| p{0.6cm}|p{1.5cm}|p{1cm}|p{1cm}|p{1cm}|p{1cm}|p{1cm}|p{1cm}|p{1cm}|p{1.5cm}|p{1.5cm}|}
 \hline
 \multicolumn{13}{|c|}{Stochastic bursts}
 \\
 \hline
 \hline
 $N_{\rm seg}$ &$N$& $T$ &$\xi$&$r_{\rm min}$ &$r_{\rm max}$ &$r_{\rm ref}$ &$\Omega_{\rm ref}$ & $\langle\Omega_b\rangle$ & $f_{\rm low}$ & $f_{\rm high}$ &  $\langle\rho_{\rm seg, stoch}\rangle$ & $\rho_{\rm tot, stoch}$\\
 \hline
 4$\times 10^4$ & 2048 & 4 s & 2.98$\times 10^{-3}$ & 2 Mpc & 5 Mpc & 2 Mpc & 2.61 & 0.803 & 20 Hz & 256 Hz & 5.04 & 3\\
 \hline
\end{tabular}
\\~\\~\\
\begin{tabular}{ |p{1cm}|p{1cm}| p{0.6cm}|p{1cm}|p{1cm}|p{1cm}|p{1cm}|p{1cm}|p{1cm}||p{1.5cm}|p{1.5cm}|p{1.5cm}|p{1.5cm}|p{1.2cm}|}
 \hline
 \multicolumn{14}{|c|}{Deterministic chirps}
 \\
 \hline
 \hline
 $N_{\rm seg}$ &$N$& $T$ & $r_{\rm min}$ &$r_{\rm max}$ & $f_{\rm low}$ & $f_{\rm high}$ & $m$ &$\langle\Omega_b\rangle$& $\xi$& $\langle\rho_{\rm seg, stoch}\rangle$ & $\rho_{\rm tot, stoch}$& $\langle\rho_{\rm seg, det}\rangle$ & $\rho_{\rm tot, det}$\\
 \hline
 4$\times 10^4$ & 2048 & 4 s & 2 Mpc & 5 Mpc & 20 Hz & 256 Hz & 30 $M_{\odot}$ & 0.803 & 2.98$\times 10^{-3}$ & 5.04 & 3 & 13.2 & 7.86\\
 \hline
\end{tabular}
\caption{Parameters used for the different analyses in Sec. \ref{s:toymodels}. Parameters listed in `Extension of previous work' and `Stochastic bursts' were used in the production of Fig. \ref{f:toymodel_2_results} and Fig. \ref{f:stoch_bursts_results}, respectively. The first 9 columns in `Deterministic chirps' were used in the production of Fig. \ref{f:DeterministiclnBFPlots}, while the last 5 columns specified the additional parameters used for Fig. \ref{f:DeterministicCornerPlot}.}
\label{t:parameter table}
\end{table}

We generate multi-sample ($N=2048$) bursts of white stochastic GWs having duty cycle $\xi=2.98\times10^{-3}$, with signal samples drawn from a probability distribution that depends on the distance $r$ to an individual source, as described above. With the chosen parameters (listed explicitly in Table \ref{t:parameter table}) the population-averaged variance 
is $\langle\sigma_b^2\rangle=0.0769$ and the noise variances are $\sigma_{n_1}^2=\sigma_{n_2}^2=0.691$. An example of the simulated data is shown 
in Fig.~\ref{f:toymodel_2_data}, together with the distribution of the burst variances $\sigma_b^2(r)$.

We analyse the data with SSC and SSI, using the full version of the likelihoods, i.e. inferring the noise parameters as well as the population parameters. We will not consider DSI for this particular data. The concrete expressions for the likelihoods can be found in Appendix \ref{app:LikelihoodsWhite}. In Fig.~\ref{f:toymodel_2_results}, we display the recovery of our SSI search, illustrating that the generalisations made in this section still allow for a successful recovery of the population and noise parameters.

We note that given the large number of samples per segment ($N=2048$) used for this analysis, one could have resorted to the reduced version of the likelihoods, where the estimates of the noise parameters are used (as provided in Appendix \ref{s:ssi_red}). We refrain from entering into a detailed comparison between full and reduced implementations of the likelihoods, as this was the topic of work by Matas and Romano \cite{Matas-Romano:2021}. Throughout the remainder of the paper, we will work with a large number of samples per segment and will employ the reduced version of the likelihoods.
\begin{figure}[hbtp!]
\begin{center}
\includegraphics[clip=true, angle=0, width=0.75\textwidth]{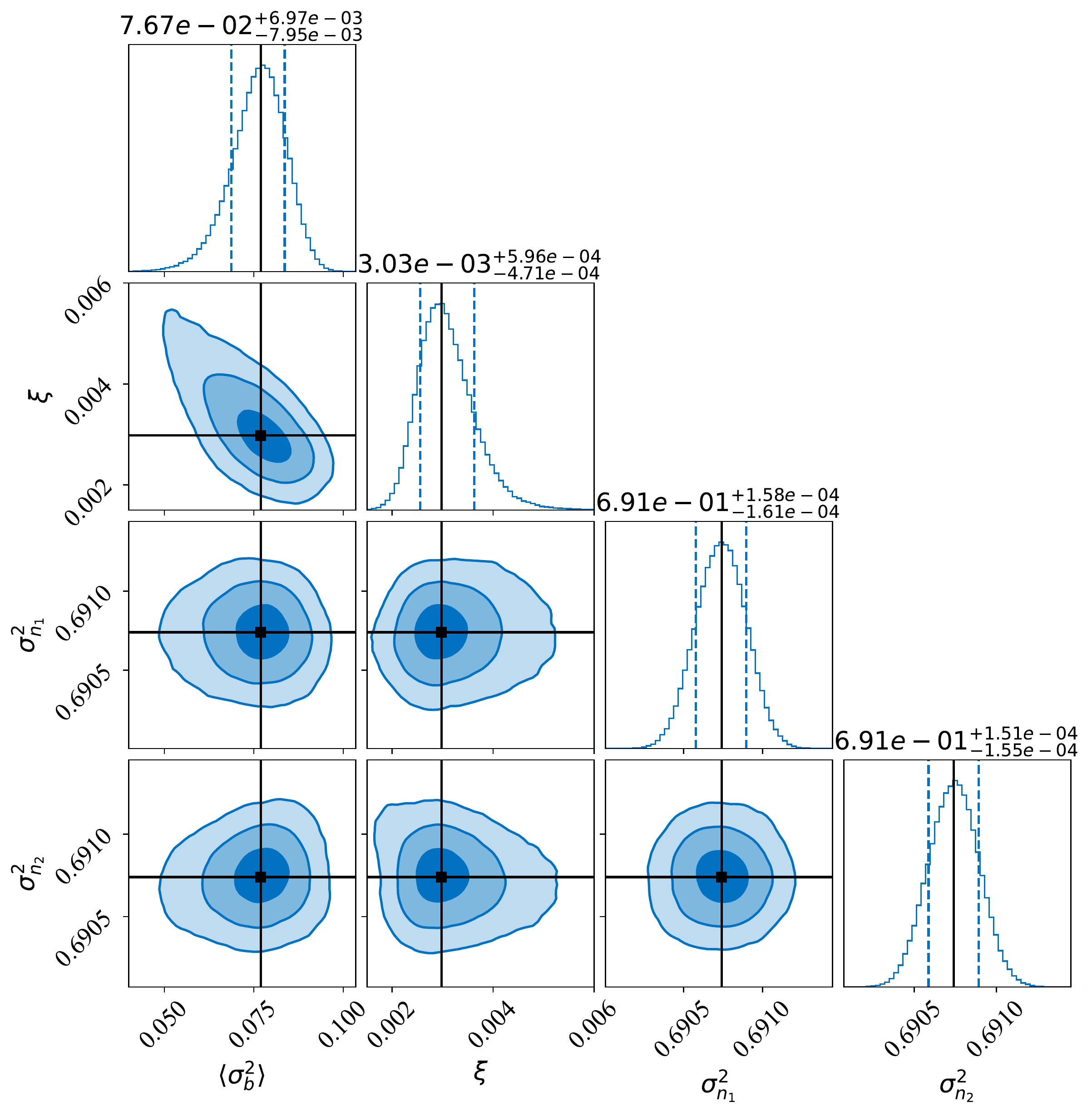}
\caption{Corner plot for the full version of the SSI analysis, combining the posteriors of 100 realizations of the data. The black lines show the injected values of the parameters used for the simulated data, and the three shaded regions for the 2-d joint posteriors correspond to 1$\sigma$, 2$\sigma$, and 3$\sigma$ uncertainty levels. All parameters are recovered within a $1\sigma$ credible interval.}
\label{f:toymodel_2_results}
\end{center}
\end{figure}


\subsection{Stochastic bursts}
\label{s:Stochastic bursts}
We extend the analysis described in the previous section to include frequency dependence. We analyze data defined by multi-sample ($N\gg 1$) bursts of stochastic 
GWs having duty cycle $\xi$ and an $f^{-7/3}$ power spectrum, for a 
uniform-in-volume distribution of source distances 
between $r_{\rm min}$ to $r_{\rm max}$, as in Section \ref{s:Previous results}.
The choice of spectral index $-7/3$ is appropriate for compact binary inspiral.
We first simulate 
data for a source at reference distance $r_{\rm ref}$ so
that it has the power spectrum%
\footnote{In practice, we first simulate the data in the
frequency domain with an amplitude spectral density 
$\sqrt{P_{\rm ref}(f)}$ and random phases, and then 
inverse-Fourier-transform the data back to the time domain.}
\be
P_{\rm ref}(f) = A_{\rm ref} \left(\frac{f}{f_{\rm ref}}\right)^{-7/3}\,,
\ee
where $A_{\rm ref}$ is some fixed amplitude, and
$f_{\rm ref}$ is a reference frequency, usually taken to be
25~Hz in line with LVK searches.
For a source at a general distance $r$, we do the same as above,
and then rescale the amplitude of the simulated signal 
by a factor of $r_{\rm ref}/r$,
which is equivalent to having
\be 
A_b(r) \equiv A_{\rm ref}\,\frac{r^2_{\rm ref}}{r^2}
\ee
as the amplitude of the power spectral density for a GW burst
at source distance $r$.
The power spectrum of a burst is therefore 
\be
P_b(r;f) = A_{\rm ref} \frac{r_{\rm ref}^2}{r^2}\left(\frac{f}{f_{\rm ref}}\right) ^{-7/3}\,.
\ee
Note that by using \eqref{eq:OmegaToPSD}, we can also write the above expression in terms of the fractional energy density spectrum $\Omega_{b}(r;f)$.
Then by taking $f=f_{\rm ref}$, we can define the amplitude of the energy density at reference frequency $f_{\rm ref}$ of a burst at distance $r$
\be
\label{e:OmegaToP2}
\Omega_b(r)\equiv\frac{10 \pi^2}{3H_0^2}f_{\rm ref}^3 P_b(r;f_{\rm ref}) = \Omega_{\rm ref}\frac{r_{\rm ref}^2}{r^2},
\qquad
\Omega_{\rm ref} \equiv \frac{10\pi^2}{3H_0^2}f_{\rm ref}^3 A_{\rm ref}.
\ee
By following the same derivation given in \eqref{e:sigma2b_popavg}, the population-averaged energy density amplitude for sources distributed uniformly-in-volume 
between $r_{\rm min}$ and $r_{\rm max}$ is 
\be
\langle\Omega_b\rangle
= 3\Omega_{\rm ref} 
\frac{r^2_{\rm ref}(r_{\rm max}-r_{\rm min})}{r_{\rm max}^3 - r_{\rm min}^3}\,.
\label{e:Omegab_popavg}
\ee
The probability distribution of the amplitude of the energy density of the bursts $\Omega_b(r)$ has the same form as \eqref{e:p_sigma2b_popavg}
\be
p(\Omega_b(r)|\langle \Omega_b\rangle)= 
\frac{\langle \Omega_b\rangle \Omega^{1/2}_{b,{\rm max}}}
{\sqrt{-3 + 12{\Omega_{b,{\rm max}}}/{\langle \Omega_b\rangle}}-3}\,
\Omega^{-5/2}_b(r)\,,
\qquad
\Omega_{\rm b,{\rm max}}\equiv \Omega_b(r_{\rm min})\,.
\label{e:p_Omegab_popavg}
\ee

Thus, the signal segment likelihood used for SSI is given by \eqref{e:likefinal_colored_nonstationary-signal} (full) and \eqref{e:SSI-reduced-colored-Zs} (reduced) with prior given by \eqref{e:p_Omegab_popavg} (i.e., $\pi(\Omega_{b,I}|\langle \Omega_b\rangle)=p(\Omega_{b}(r_I)|\langle \Omega_b\rangle)$). The integration bounds are then $\Omega_{\rm b, min}(\langle\Omega_b\rangle)$ and $\Omega_{\rm b, max}$ where 
$\Omega_{\rm b, min} = \Omega_{b}(r_{\rm max})$ and $r_{\rm max}$ is written in terms of the population parameter $\langle\Omega_b\rangle$, in the same manner as \eqref{e:rmax_sigma2b}.

For reference, we note that the expected value 
of the stochastic (optimally-filtered) signal-to-noise 
ratio for a segment that contains a GWB burst is
\be
\rho_{\rm seg, stoch} = \sqrt{2 T}
\left[\int_{f_{\rm low}}^{f_{\rm high}} {\rm d}f\>
\frac{P_b^2(f)}{P_{n_1} P_{n_2}}\right]^{1/2}\,,
\label{e:rho_seg,stoch}
\ee
where $P_{n_1}$ and $P_{n_2}$ are the power spectra of the noise in each detector. 
Note that, if the two detectors were not co-located and co-aligned,
we would need to include a factor of the overlap reduction squared 
in the numerator of the integrand in \eqref{e:rho_seg,stoch}.
The above expression for $\rho_{\rm seg, stoch}$ is a
{\em power} signal-to-noise ratio, defined as the expected
value of the optimally-filtered cross-correlation statistic 
divided by its standard deviation, 
see, e.g., \cite{Romano-Cornish:2017}.

As mentioned before, our stochastic-signal-based search looks for a GWB consistent
with a power spectrum of spectral index $-7/3$, as expected for BBH mergers. In 
contrast, the deterministic-signal-based search described in Sec. \ref{s:DSI} (which we 
call DSI) looks for deterministic BBH chirp waveforms, where the signal parameters 
of the individual chirps must be marginalized over. 
We inject intermittent, stochastic bursts with an $f^{-7/3}$ power spectrum and duty cycle $\xi=2.98\times10^{-3}$. The parameters used for the injection are displayed in Table \ref{t:parameter table}.We arbitrarily choose the reference distance $r_{\rm ref} = r_{\rm min}$. The value of $\Omega_{\rm ref}$ is chosen to be $2.61$ (to be consistent with the parameters chosen in Sec. \ref{s:Deterministic bursts}). With these parameters, the population-averaged energy density amplitude of the bursts is $\langle\Omega_b\rangle=0.803$. The noise is then set such that the average SNR per segment, as computed with \eqref{e:rho_seg,stoch}, is 5.04 to give a total SNR of 3, as obtained by using \eqref{e:rho_seg}.

\begin{figure}[hbtp!]
\begin{center}
\includegraphics[clip=true, angle=0, width=0.48\textwidth]{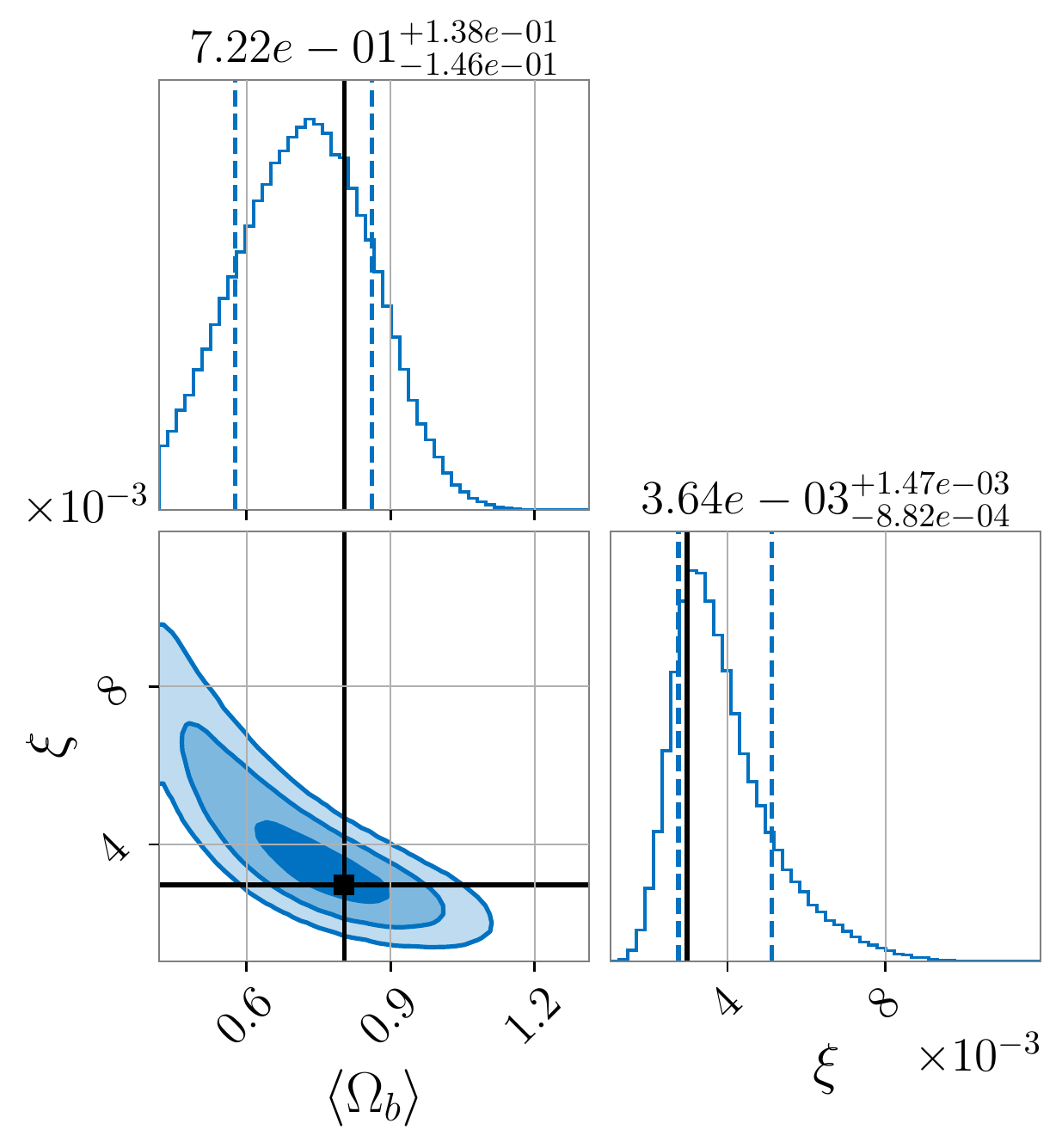}
\includegraphics[clip=true, angle=0, width=0.49\textwidth]{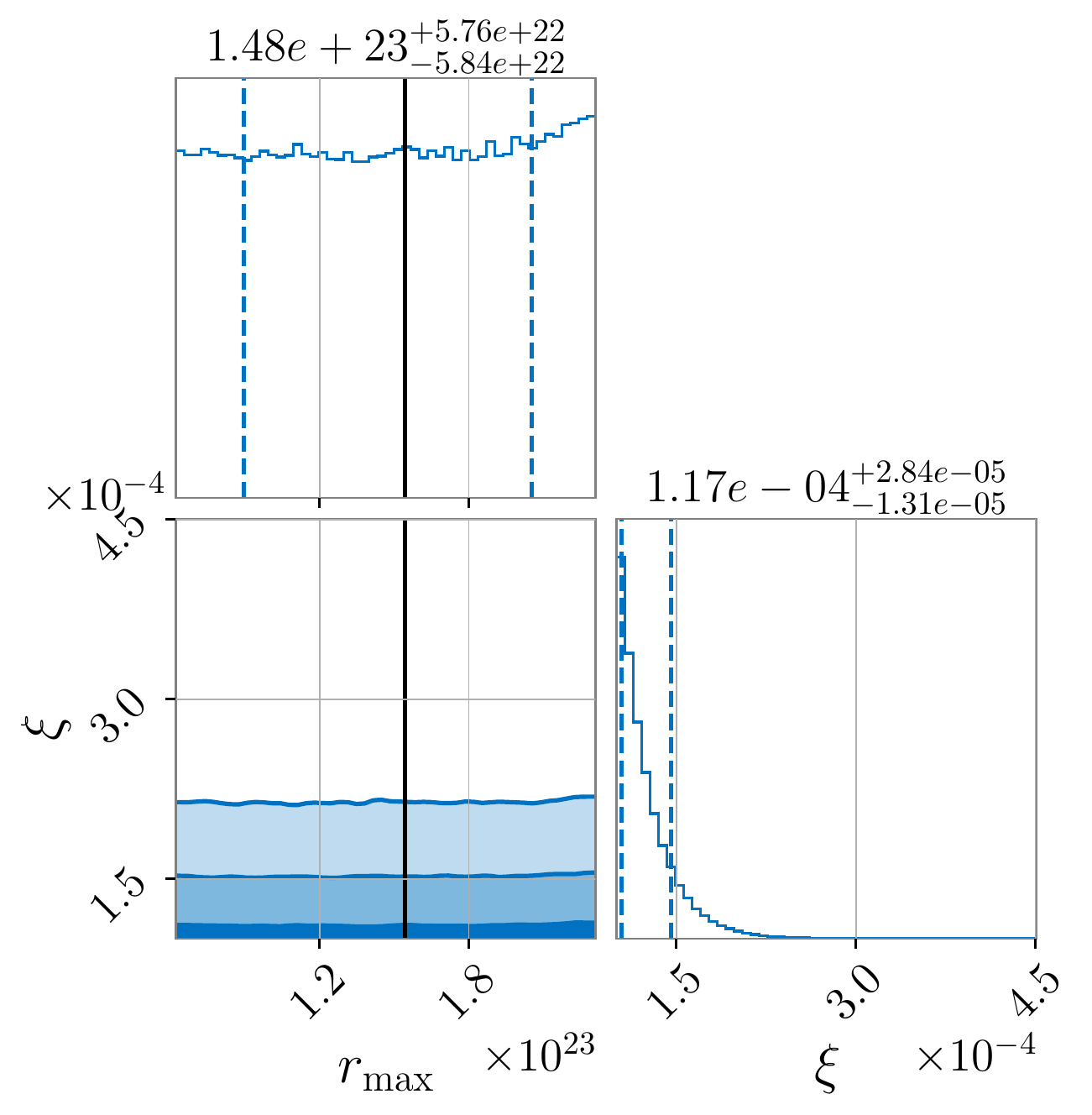}
\caption{For intermittent, stochastic bursts with an $f^{-7/3}$ power spectrum, we demonstrate recovery of our search (left) and compare it to that of a deterministic-signal-based search (right). Our search recovers the injected signal parameters within a $1\sigma$ credible interval, while DSI recovers the uniform prior on $r_{\rm max}$ and the lower boundary of the prior imposed on the duty cycle ($\xi=10^{-4}$). Thus, the DSI analysis finds no signal in the data.
}
\label{f:stoch_bursts_results}
\end{center}
\end{figure}

We analyze our data with the reduced forms which estimate the noise parameters of our stochastic-signal-based search (SSI) and the deterministic-signal-based search (DSI). The exact form of the likelihood is given in Sec. \ref{s:SSI} (with coarse-graining factor $M=16$) and \ref{app:DSIReducedColored}, respectively. The population parameter recovered by SSI is $\langle\Omega_b\rangle$ while the population parameter recovered by DSI is $r_{\rm max}$. Note, these are related by \eqref{e:Omegab_popavg}. In Fig.~\ref{f:stoch_bursts_results}, we demonstrate that DSI cannot recover the signal in the data, since no chirp waveform exists. While this result is in a sense obvious, it highlights the challenges that a deterministic-signal-based search faces. Incorrectly modeling the waveforms of the chirps could lead the search to overlook a signal which is present. Conversely, SSI recovers both stochastic bursts of GW power as well as deterministic waveforms, as we will see in the next section.

\subsection{Deterministic chirps}
\label{s:Deterministic bursts}

Finally, we consider multi-sample bursts of GWs produced by 
deterministic BBH chirp signals, for
a uniform-in-volume distribution of sources \eqref{e:unif-in-vol}.
The corresponding power spectrum will necessarily have an approximate 
$f^{-7/3}$ frequency dependence.
By using deterministic BBH chirp signals, this analysis is 
more in line with the assumptions
made by the deterministic-signal-based search DSI.

We assume that all parameters 
defining the chirp waveforms except for the distances
to the sources (e.g., the chirp mass 
${\cal M}_c \equiv (m_1 m_2)^{3/5}/(m_1+m_2)^{1/5}$,
the inclination angle $\iota$, the coalescence time $t_{\rm col}$,
and the phase of coalescence within a segment) 
have fixed values and are known 
a~priori by the DSI search.
For simplicity, we choose the two component masses to be
equal (i.e., $m_1=m_2\equiv m$);
the inclination angle $\iota=\pi/2$ so that the source is linearly polarized
(i.e., $h(t) = h_+(t)$, $h_\times(t)=0$); the phase at coalescence
$\Phi_0$ to be zero; and the coalescence
time $t_{\rm col}$ to occur at the end of a segment, so $t_{\rm col}=T$, the segment duration.
For a source drawn from the population with distance $r$, 
the explicit form for the 
simulated deterministic chirp signal is given in the time domain by \cite{Maggiore-book}
\be
h_{\rm chirp}(t; r)=
\frac{1}{2r}
\left(\frac{G{\cal M}_c}{c^2}\right)^{5/4}
\left(\frac{5}{c\tau}\right)^{1/4}
\cos\left[\Phi(\tau)\right]\,,
\qquad
\tau\equiv t_{\rm col}-t\,,
\label{e:h_model_chirp}
\ee
where
\be
\Phi(\tau)
\equiv -2\left(\frac{5 G{\cal M}_c}{c^3}\right)^{-5/8} \tau^{5/8} +\Phi_0
\ee
encodes the frequency evolution of the chirp,
\be
f(t) \equiv 
-\frac{1}{2\pi}\frac{{\rm d}}{{\rm d}\tau}\Phi(\tau)
= \frac{1}{\pi}
\left(\frac{G{\cal M}_c}{c^3}\right)^{-5/8}
\left(\frac{5}{256}\frac{1}{\tau}\right)^{3/8}\,.
\ee
The corresponding BBH chirp power spectrum is
\be
P_{\rm chirp}(r;f) = \frac{2}{T} \left|\tilde h_{\rm chirp}(r;f)\right|^2
\equiv 
A_{\rm chirp}(r)\left(\frac{f}{f_{\rm ref}}\right)^{-7/3}\,,
\ee
where $\tilde h_{\rm chirp}$ is the Fourier transform of the chirp waveform and
\be
A_{\rm chirp}(r) = 
A_{\rm ref}\frac{r_{\rm ref}^2}{r^2}\,,
\qquad
A_{\rm ref} \equiv
\frac{2}{T}
\frac{c^2}{4r_{\rm ref}^2}
\left(\frac{5\pi}{24}\right)
\left(\frac{G{\cal M}_c}{c^3}\right)^{5/3}(\pi f_{\rm ref})^{-7/3}\,.
\ee
Note one can express the chirp PSD, $P_{\rm chirp}$, in terms of the fractional energy density of the chirps by using \eqref{eq:OmegaToPSD}. For reference, we note that the expected value 
of the deterministic (matched-filter) signal-to-noise ratio 
for a segment which contains a BBH chirp signal is \cite{Romano-Cornish:2017}
\be
\rho_{\rm seg, det} 
=\left[4 \sum_{\mu=1}^2
\int_{f_{\rm low}}^{f_{\rm high}} {\rm d}f\>
\frac{|\tilde h_{\rm chirp}(f)|^2}{P_{n_\mu}}\right]^{1/2}
=\sqrt{2 T}
\left[\sum_{\mu=1}^2 
\int_{f_{\rm low}}^{f_{\rm high}} {\rm d}f\>
\frac{P_{\rm chirp}(f)}{P_{n_\mu}}\right]^{1/2}\,,
\label{e:rho_seg,det}
\ee
where $P_{n_\mu}$ is the noise power spectral density
in detector $\mu=1,2$ (see \eqref{e:noise_power}).
The above expression for $\rho_{\rm seg, det}$ is an
{\em amplitude} signal-to-noise ratio, defined as the 
expected value of the matched-filter statistic 
divided by its standard deviation.
The quadrature sum takes into account the contribution
from using both detectors to do the analysis.

\begin{figure}[hbtp!]
\begin{center}
\includegraphics[clip=true, angle=0, width=0.48\textwidth]{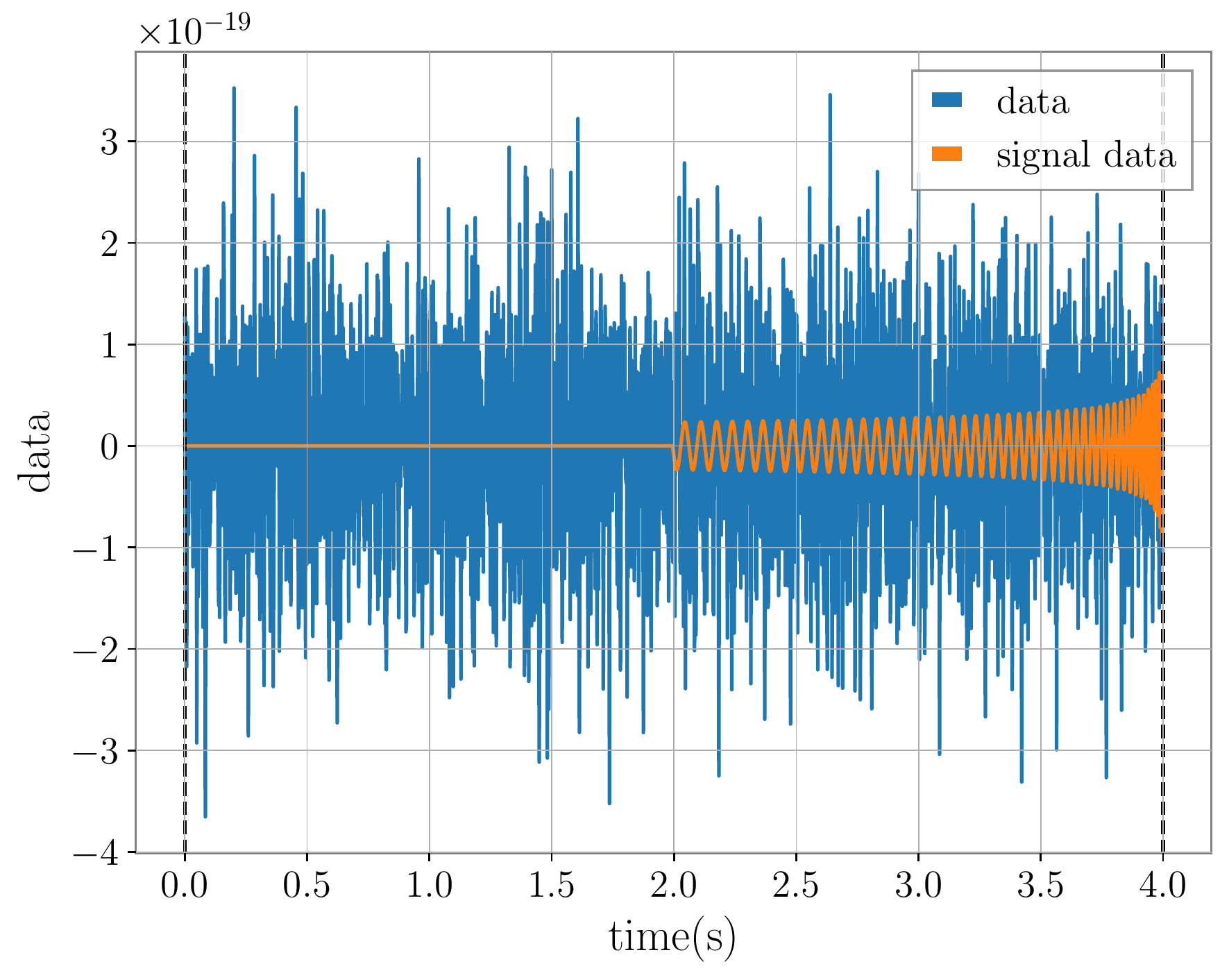}
\includegraphics[clip=true, angle=0, width=0.49\textwidth]{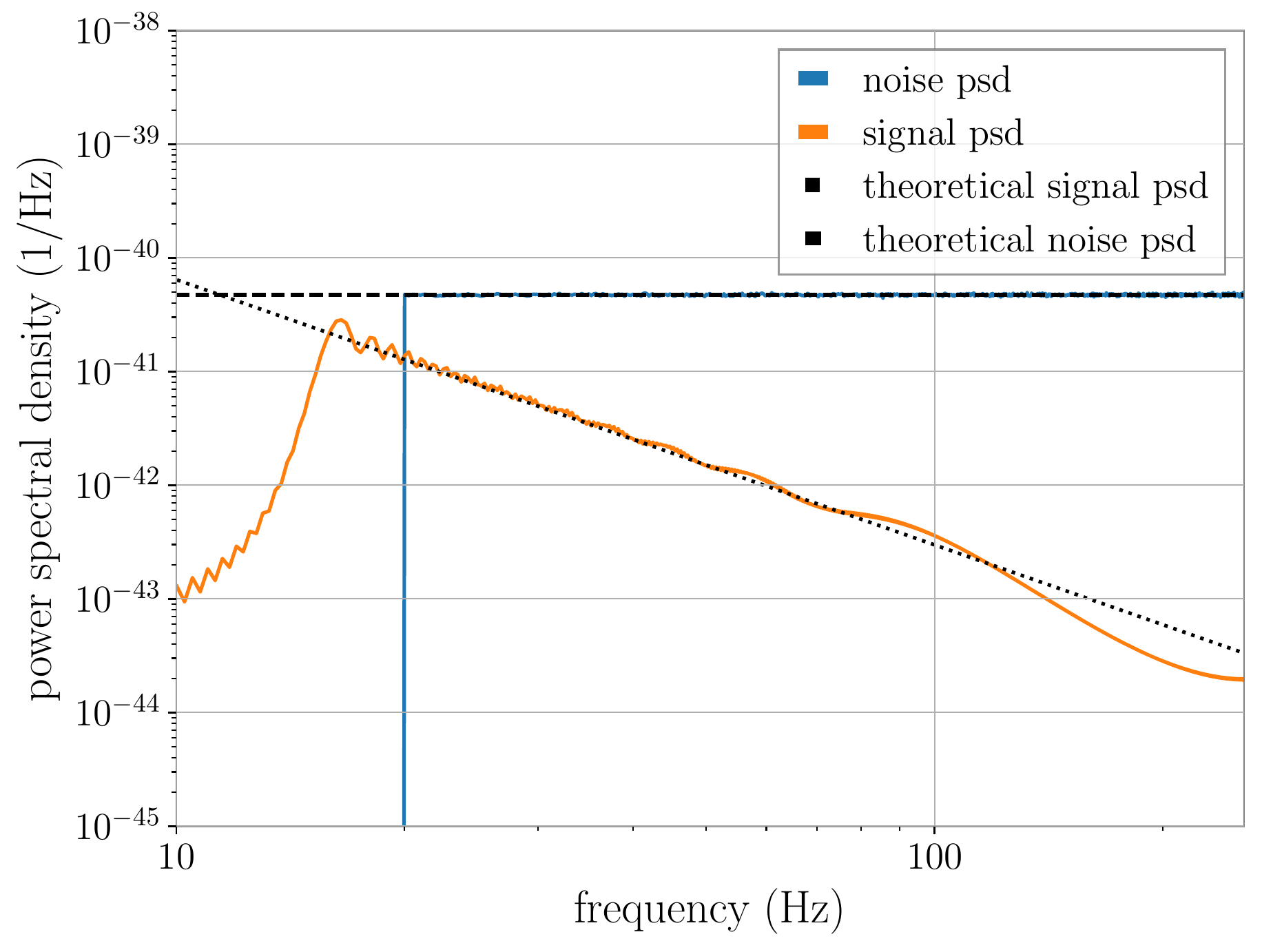}
\caption{\textbf{Left:} Example BBH chirp signal in the time-domain as given by \eqref{e:h_model_chirp}. \textbf{Right:} Averaged power spectral density of an ensemble of BBH chirp signals as a function of frequency for the noise and signal separately, together with their theoretical predictions according to the injected values.}
\label{f:bbh-chirp}
\end{center}
\end{figure}

Figure~\ref{f:bbh-chirp} shows a plot of a representative BBH chirp 
signal in the time-domain (left panel) and an average over an ensemble of BBH chirp signals in the frequency domain (right panel). 

As mentioned in Section \ref{s:DF}, the detection statistic in our Bayesian framework is the Bayes factor where the models in \eqref{e:bayes_factor} are the signal+noise model and the noise only model for a particular search. While SSC and SSI contain the same noise model, the noise model in DSI does not take the same form. Hence, the Bayes factors for the different searches are not computed with respect to the same noise model and one cannot compare these methods with one another in terms of the Bayes factor. Instead, we evaluate how the intermittent nature of the signal impacts each search method's effectiveness in recovering the signal by plotting the ln Bayes factor as a function of the duty cycle. In other words, we wish to answer two questions: (i) How well does SSI do in recovering the signal at different duty cycles for a constant total stochastic signal-to-noise ratio? and (ii) How well does DSI do in recovering the signal at different duty cycles for a constant total deterministic signal-to-noise ratio?
The answers to the questions are independent of one another and cannot be used as a way to assess if one search is ``better'' than the other. However, since SSC and SSI contain the same noise model, these searches can be compared to one another using the Bayes factor.

In order to assess the efficiency of the methods with respect to their respective noise-only models, we simulate 40,000 segments of data with each segment being 4 seconds long. We choose values of $r_{\rm min}=2$ Mpc, $r_{\rm max}=5$ Mpc and the black hole component masses to each be $30 M_{\odot}$. These parameters give a value of $\langle\Omega_b\rangle=0.803$. The parameters used for this analysis are tabulated in the first 9 columns of the `Deterministic chirps' section in Table \ref{t:parameter table}. Thus, the signal has the same strength as in \ref{s:Stochastic bursts}, but it is now composed of deterministic chirps. The same coarse-graining factor and low- and high-frequency cutoffs that were used in Section \ref{s:Stochastic bursts} are used for this case as well when analyzing the data.

\begin{figure}[hbtp!]
\begin{center}
\includegraphics[clip=true, angle=0, width=0.46\textwidth]{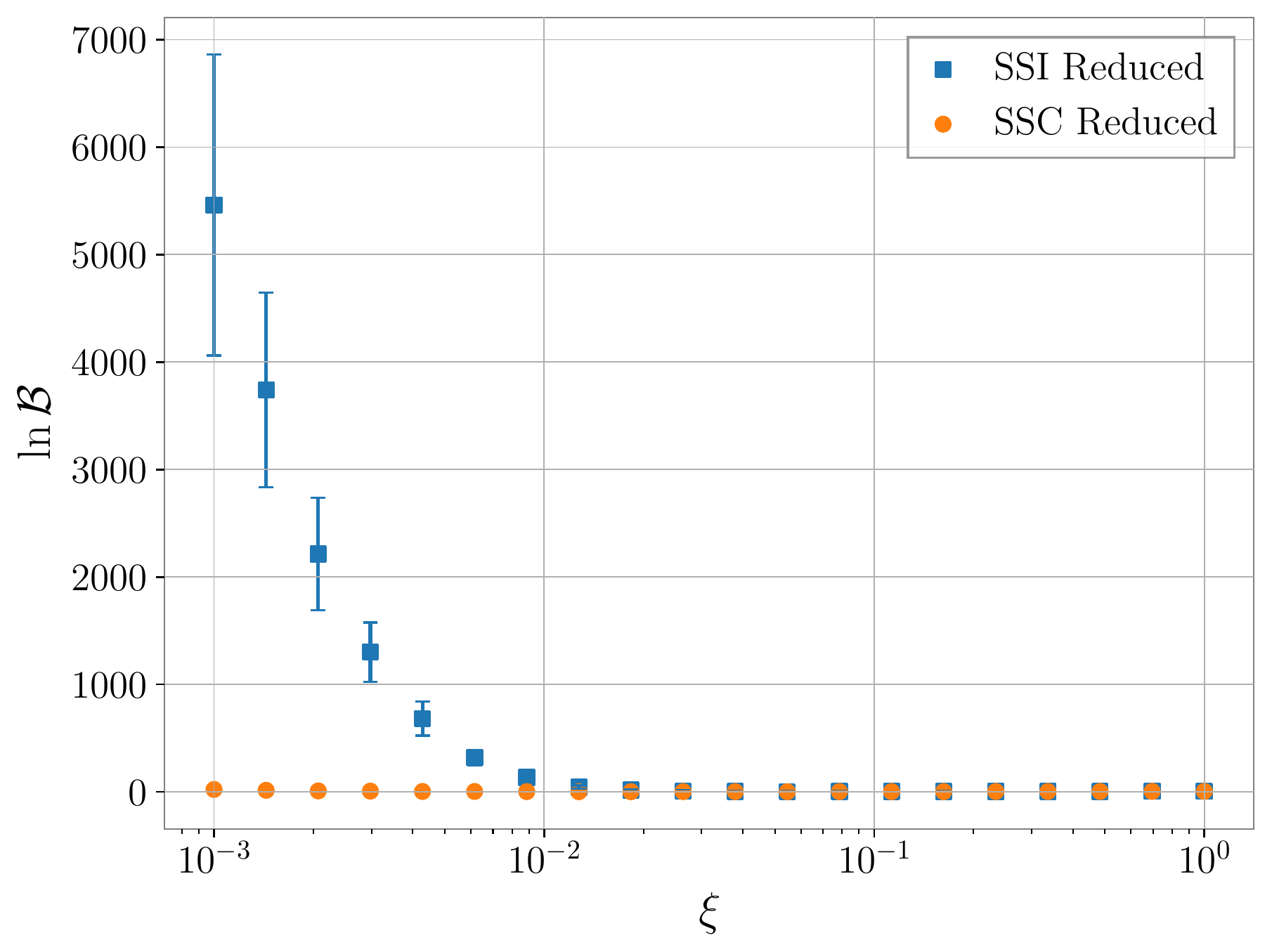}
\includegraphics[clip=true, angle=0, width=0.46\textwidth]{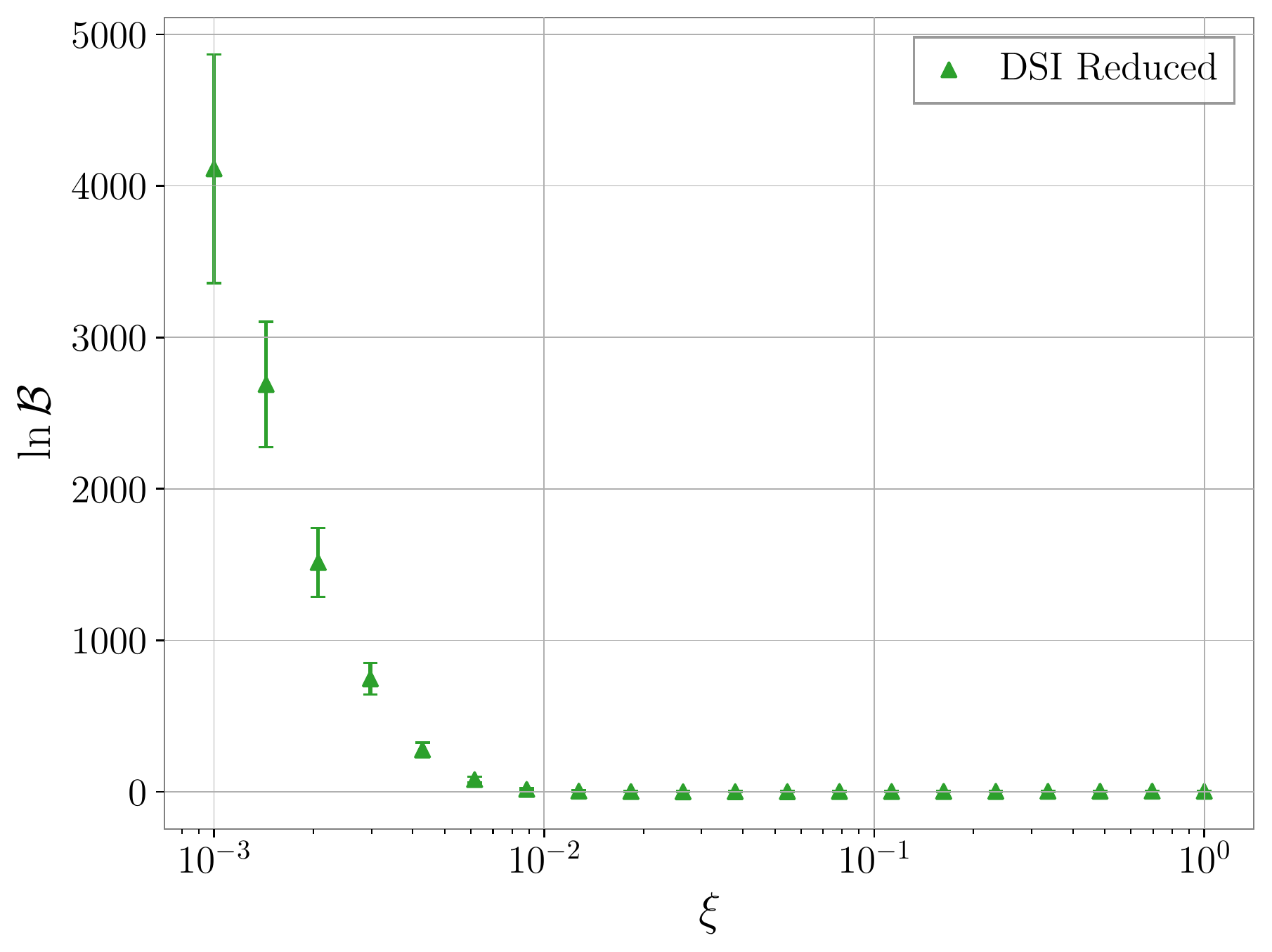}
\caption{Plots of the ln Bayes factor averaged over 100 data realizations for SSC and SSI (left) and DSI (right) for deterministic chirp signals occurring with various values of the duty cycle $\xi$. Both intermittent searches are well-suited for detecting signals with a low duty cycle.}
\label{f:DeterministiclnBFPlots}
\end{center}
\end{figure}

Figure~\ref{f:DeterministiclnBFPlots} shows the ln Bayes factors for the stochastic-signal-based searches (left panel) and for the deterministic-signal-based search (right panel) as a function of the duty cycle. Analogously to what was done in Section \ref{s:DF}, the total SNR is kept constant by adjusting the noise levels. For the stochastic searches, we keep the total power SNR, computed using \eqref{e:rho_seg,stoch}, constant, while for the deterministic search we keep the total amplitude SNR constant, obtained using \eqref{e:rho_seg,det}. 
We see that both intermittent searches (SSI and DSI) perform well at low duty cycles, with values of the ln Bayes factors reaching over 1000 for some of the smallest values of the duty cycle considered.

\begin{figure}[hbtp!]
\begin{center}
\includegraphics[clip=true, angle=0, width=0.95\textwidth]{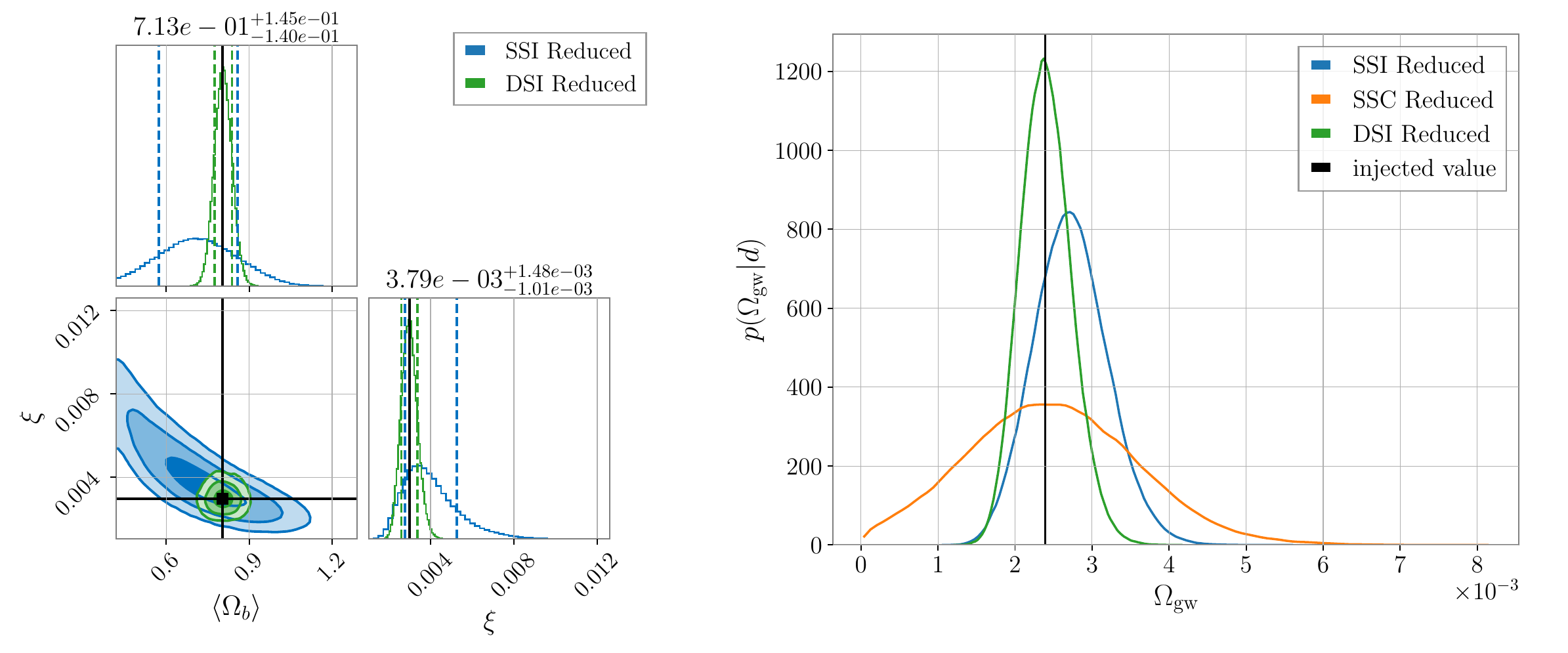}
\caption{\textbf{Left:} Posterior corner plot combined over 100 data realizations analyzed with SSI Reduced (blue) and DSI Reduced (green). Both searches recover the injected signal parameters ($\xi=2.98\times10^{-3}$ and $\langle\Omega_b\rangle=0.803$) within a 1$\sigma$ confidence interval. The recovered values and error bars are those recovered by the SSI Reduced search. \textbf{Right: } 1D posterior plot of $\Omega_{\rm gw}$ samples from SSI Reduced (blue), SSC Reduced (orange) and DSI Reduced (green) constructed by combining posterior samples for $\xi$ and $\langle\Omega_{b}\rangle$ using \eqref{e:Omega_gw}.
Note, the inference done with the DSI likelihood gives posterior samples for the parameters $\xi$ and $r_{\rm max}$ and the values of $r_{\rm max}$ are then converted to samples in $\langle\Omega_b\rangle$ by \eqref{e:Omegab_popavg}, since the other variables in \eqref{e:Omegab_popavg} are fixed and known.
}
\label{f:DeterministicCornerPlot}
\end{center}
\end{figure}

In order to directly compare SSI with DSI, we run both analyses on the same dataset. The data is generated such that the duty cycle is $2.98\times10^{-3}$, the signal is the same as described above and the noise variance is chosen such that the average stochastic SNR per segment, computed using \eqref{e:rho_seg,stoch}, is equal to $5.04$ and the total stochastic SNR is equal to $3.0$. Note for these values, the average deterministic SNR per segment, computed using \eqref{e:rho_seg,det}, is $13.20$ with the total deterministic SNR being $7.86$, which is considerably larger than the total stochastic SNR. Note these parameters are displayed in the remaining columns of the `Deterministic chirps' section of Table \ref{t:parameter table}. A comparison of the recovered corner plots is shown in Fig.~\ref{f:DeterministicCornerPlot} (left panel). We see that for this data, both searches recover the signal within a 1$\sigma$ credible interval, with the error bars for DSI much smaller than SSI, due to the deterministic approach appropriately modeling the chirp waveform of the signal. We also show a comparison of 1D posterior plots of $\Omega_{\rm gw}$ in Fig.~\ref{f:DeterministicCornerPlot} (right panel). Similarly to the corner plot, the posterior width is smaller for DSI than SSI, although SSI still performs better than SSC.

\begin{figure}[hbtp!]
\begin{center}
\includegraphics[clip=true, angle=0, width=0.48\textwidth]{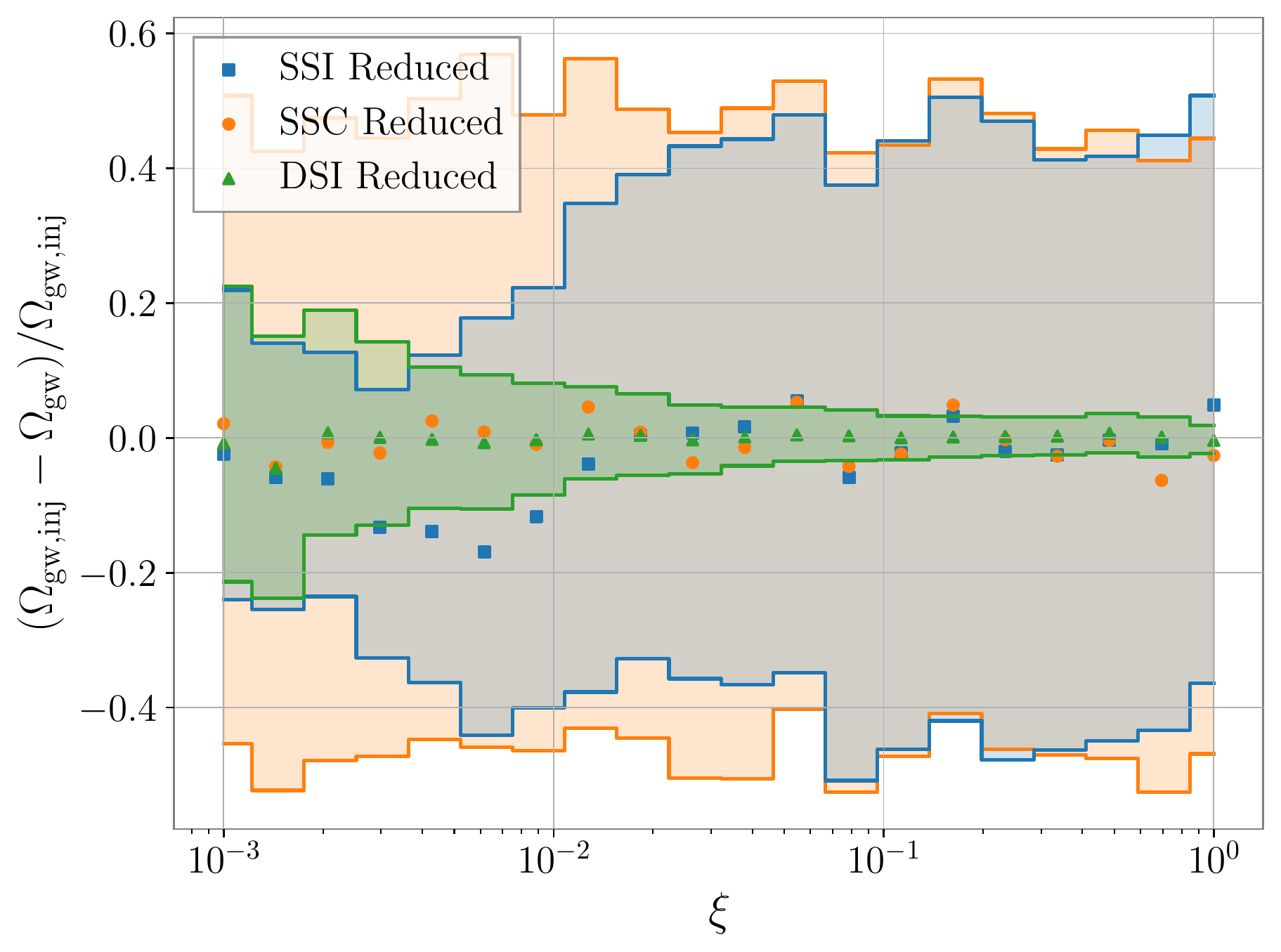}
\includegraphics[clip=true, angle=0, width=0.46\textwidth]{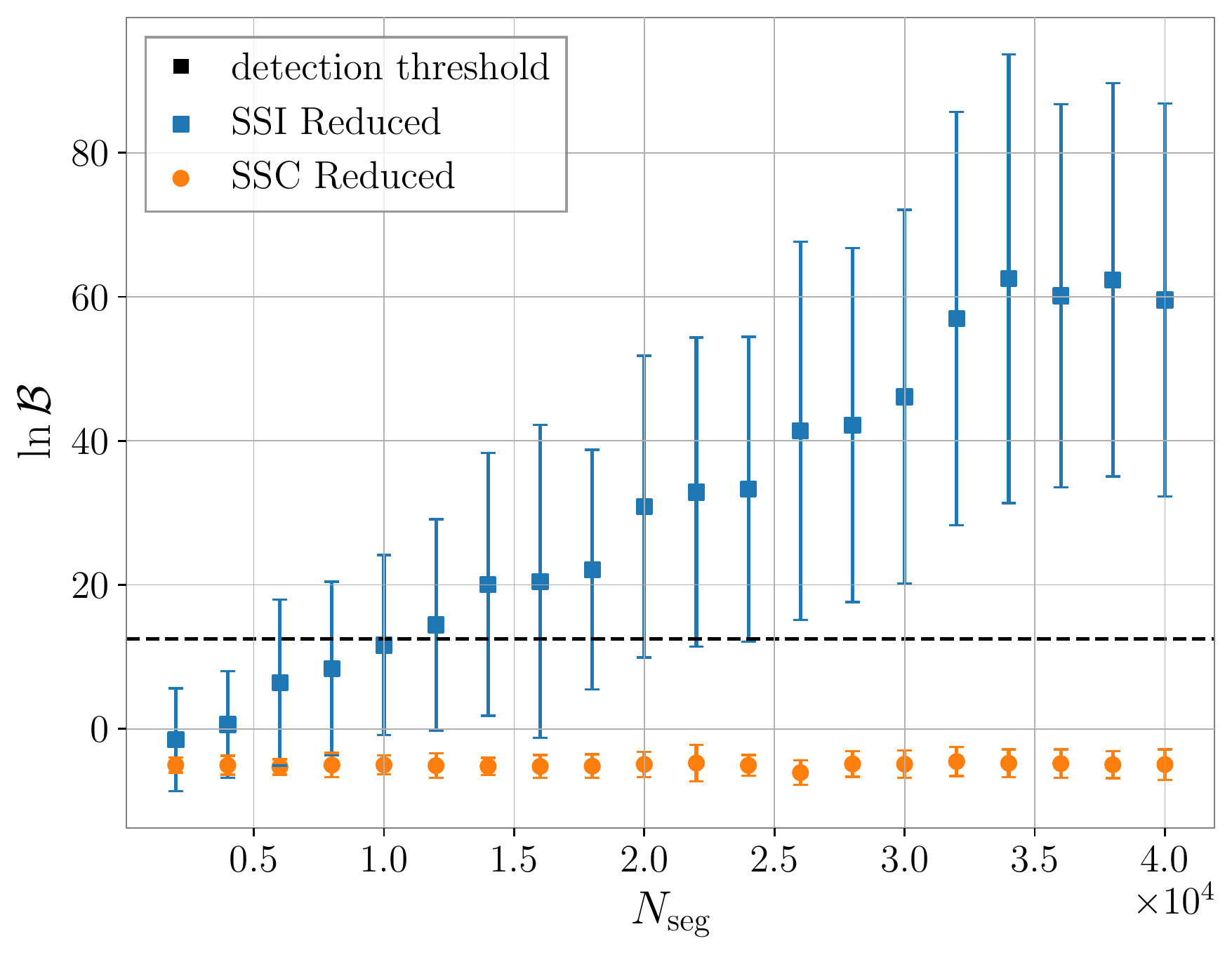}
\caption{\textbf{Left:} Comparison of recovered values to injected value of $\Omega_{\rm gw}$ for SSI Reduced (blue), SSC Reduced (orange) and DSI Reduced (green) for different values of the duty cycle. All injected parameters are equivalent to the parameters used in the left panel of Fig. \ref{f:DeterministiclnBFPlots} and the recovered values are those after combining 100 realizations of data. The shaded regions represent the 1$\sigma$ credible interval of the combined 100 realizations of data. \textbf{Right:} ln Bayes factor vs $N_{\rm seg}$ for data with $\langle\rho_{\rm seg, stoch}\rangle = 2$ and $\xi=2.98\times10^{-3}$. We define a detection threshold of $\ln \pazocal{B}=12.5$. SSI crosses this threshold after $\sim 12,000$ segments of data, while SSC crosses this threshold after $\sim 650,000$ segments of data, corresponding to a factor of improvement in detecting the signal of roughly 54 for SSI relative to SSC.} 
\label{f:error_lnBFvsNSeg}
\end{center}
\end{figure}

One notes a small bias in the recovery of $\Omega_{\rm gw}$ for SSI in the right panel of Fig.~\ref{f:DeterministicCornerPlot}. In Fig.~\ref{f:error_lnBFvsNSeg} we show the relative difference of the injected value and recovered value of $\Omega_{\rm gw}$ as a function of $\xi$ for the three searches, together with the 1$\sigma$ uncertainty band, after combining the posterior over 100 realizations of data. We note that the biased recovery is not always towards higher values of $\Omega_{\rm gw}$. We also note that the width of the uncertainty for the DSI analysis improves as $\xi$ increases because the total deterministic SNR is not held constant and increases.

To conclude, we give an estimate of the improvement in time to detection of a GWB with our search. Note that this estimate is computed under the assumptions adopted in this paper and will therefore most likely differ for a realistic detector configuration, with realistic detector noise. We also note that the strength of the signal may affect these values. Nevertheless, to obtain such an estimate, we simulate a GWB consisting of deterministic chirps with parameters $\langle\rho_{\rm seg, stoch}\rangle=2$ (corresponding to $\langle\rho_{\rm seg, det}\rangle=8.3$) and $\xi=2.98\times 10^{-3}$. We then vary the number of data segments and assess how many 4 second segments are needed to reach a threshold value of the ln Bayes factor which is large enough to claim a detection. We define this threshold to be of value 12.5, corresponding to a detection of SNR equal to 5. This is shown in the right panel of Fig.~\ref{f:error_lnBFvsNSeg} for SSI and SSC.
Due to the large difference in deterministic and stochastic SNR, the ln Bayes factor for DSI already reaches $\sim 160$ at the first value of $N_{\rm seg}$ considered. We therefore do not include DSI on this plot to avoid scaling issues. We estimate that the SSC search would cross this threshold after 650,000 segments of data. This corresponds to a factor of $\sim 54$ improvement in detection of SSI versus SSC for these parameters and assumptions. 

\section{Discussion}
\label{s:discussion}
Developing data-analysis techniques to reduce the time-to-detection of an astrophysical GWB with the LVK detectors is one of the current challenges that the GW community faces. Searches that include the intermittency of the BBH background to improve detection statistics have been proposed in the past \cite{Drasco-Flanagan:2003,Smith-Thrane:2018, Yamamoto:2022, Buscicchio:2022}. In this work, we propose a new, stochastic search for intermittent GWBs and compare its efficiency with other searches. Our stochastic-based search looks for excess cross-correlated power in short stretches of data, ignoring the deterministic form of the GW signal waveforms and, hence the need to marginalize over all the associated signal 
parameters, as is done in the deterministic-signal-based approach of Smith and Thrane~\cite{Smith-Thrane:2018}. Not only is it beneficial to develop multiple searches in order to cross-check a potential detection, but there is an added benefit to running a search which does not look for a specific waveform in the data. The stochastic signal model allows our search to be flexible with respect to the type of signal it can detect. By changing the spectral index $\alpha$ in the search (or by allowing $\alpha$ to be inferred as a population parameter) we could detect other intermittent signals which might exist in the data.

For a series of analyses on data of increasing complexity, we show that for data with low duty cycles our search performs better than the standard continuous cross-correlation search, which does not take the intermittent nature of the BBH background into account. Furthermore, we show that a stochastic search for intermittent GWBs is more flexible to the source of the intermittent GWB than our implementation of the Smith and Thrane approach~\cite{Smith-Thrane:2018} and should be more computationally efficient in detecting a signal. The detection of an intermittent background will allow us to test existing theoretical models, as described in \cite{Mukherjee:2019oma, Mukherjee:2020jxa}.

Before being able to apply this search method on real GW data, further generalizations need to be made. We give several examples of such generalizations, which will be addressed in future work.

For all of our data in this paper, we only simulate signals which lie completely within the segment boundaries. A crucial next step is investigating how a signal which extends past a segment boundary will impact our results. 
Further, the most realistic data we consider consists of individual BBH chirps injected in white, Gaussian noise. However, various assumptions were made about the source distribution that generates these chirps. For example, the two component masses were chosen to be equal, and the resulting chirp mass chosen to be identical for all the chirps (with only the distance to the source varying from one data segment to another). In reality, the black hole masses will most likely follow a power-law + peak distribution as shown by the latest LVK results \cite{GWTC-3}. Generalizing our method to allow for such mass distributions, as well as the performance of our search in that case, is left for future work.

Several simplifications regarding the detectors were made as well. First, we worked under the assumption that the detectors are co-located and co-aligned. This needs to be generalized by taking into account the effect of the overlap reduction function. Second, it was assumed that the noise in the detector is white and Gaussian. However, realistic detector noise follows a colored, i.e. frequency-dependent, power spectral density. An additional complication related to noise estimation arises from the presence of a continuous GWB of BNS mergers. At any time, several BNS mergers are expected to be emitting GWs in the LVK frequency band. Not only does this violate the assumption that a segment contains either one signal or noise only, but it will also affect the noise PSD estimation. Challenges related to the correct noise estimation will be addressed in future work. Furthermore, the Gaussian noise assumption will likely be violated as well, due to the presence of noise transients, so-called glitches. During the third observing run of the LVK collaboration, these glitches were omnipresent in the data \cite{O3-isotropic, Davis:2021}. Therefore, before analyzing real detector data, the sensitivity of our search to the presence of such glitches will have to be investigated. Analyzing real detector data will introduce many challenges, which we plan to address incrementally, considering more and more realistic detectors and signals.

\section*{Acknowledgement}
Joseph Romano and Jessica Lawrence are supported by National Science Foundation (NSF) Grant No. PHY-2207270. Joseph Romano was also supported by start-up funds provided by Texas Tech University. Kevin Turbang is supported by FWO-Vlaanderen through grant number 1179522N. Arianna Renzini is supported by the NSF award 1912594. The authors are grateful for computational resources provided by the LIGO Laboratory and supported by NSF Grants PHY-0757058 and PHY-0823459. The Bayesian inference was performed using {\tt bilby} \cite{Ashton_2019} with the {\tt dynesty} sampler \cite{Speagle_2020}.

\appendix
\section{Likelihoods}
\label{app:Likelihoods}

Throughout this work, various searches for GWBs are compared. In this appendix, we provide the likelihoods corresponding to those searches. We start by giving an overview of the likelihoods used in Section \ref{s:Previous results}, i.e., applicable to white signals, and conclude with the likelihoods for colored signals used in Sections \ref{s:Stochastic bursts} and \ref{s:Deterministic bursts}. We also remind the reader that all likelihoods considered in this work are for stationary, white-Gaussian noise (see \eqref{e:noise_power}).

\subsection{Likelihoods for white signals}
\label{app:LikelihoodsWhite}

\subsubsection{SSC-full}

For white signals, we define the likelihood functions for a continuous stochastic search (SSC-full) 
as~\cite{Matas-Romano:2021}:
\begin{multline}
{\cal L}(d|\sigma_{\rm gw}^2,\sigma^2_{n_1},\sigma^2_{n_2})
\\=\prod_{I=1}^{N_{\rm seg}}
\frac{1}{(2\pi)^N 
\left(\sigma^2_1\sigma^2_2-(\sigma^2_{\rm gw})^2\right)^{N/2}}
\exp\left\{-\frac{1}{2}
\frac{N}{
\left(\sigma^2_1\sigma^2_2-(\sigma^2_{\rm gw})^2\right)}
\left[\hat \sigma^2_{1,I}\sigma^2_2
+ \hat \sigma^2_{2,I}\sigma^2_1
- 2 \hat\sigma^2_{{\rm gw},I} \sigma^2_{\rm gw}\right]\right\}\,,
\label{e:SSS-full-white}
\end{multline}
where
\be
\sigma^2_1\equiv \sigma^2_{n_1} + \sigma^2_{\rm gw}\,,
\qquad
\sigma^2_2\equiv \sigma^2_{n_2} + \sigma^2_{\rm gw}\,,
\ee
are parameters describing the total auto-correlated power in 
detectors 1 and 2, and
\be
\hat\sigma^2_{1,I} \equiv \frac{1}{N}\sum_{i}d^2_{1,Ii}\,,
\qquad
\hat\sigma^2_{2,I} \equiv \frac{1}{N}\sum_{i}d^2_{2,Ii}\,,
\qquad
\hat\sigma^2_{{\rm gw},I} \equiv \frac{1}{N}\sum_{i}d_{1,Ii}d_{2,Ii}\,,
\ee
are the quadratic combinations of the data from segment $I$ that
enter the likelihood function.
(Here, $i$ labels the time sample in
data segment $I$.) The noise variances in each detector are $\sigma_{n_1}^2$ and $\sigma_{n_2}^2$.
It turns out that $\hat\sigma^2_{1,I}$, $\hat\sigma^2_{2,I}$, $\hat\sigma^2_{{\rm gw},I}$
are the maximum-likelihood estimates of $\sigma^2_1$, $\sigma^2_2$, $\sigma^2_{\rm gw}$
for segment $I$.

\subsubsection{SSC-reduced}
\label{s:sss_red}
For a large number of samples per segment ($N\gg 1$), one can define a reduced version of the likelihood function, which is 
given by~\cite{Matas-Romano:2021}:
\be
{\cal L}(d|\sigma^2_{\rm gw},\bar\sigma^2_{n_1},\bar\sigma^2_{n_2})
= \prod_{I=1}^{N_{\rm seg}}
\frac{1}{\sqrt{2\pi\,{\rm var}(\bar\sigma^2_{\rm gw})}}
\exp\left[-\frac{(\hat \sigma^2_{{\rm gw},I}-\sigma^2_{\rm gw})^2}
{2\,{\rm var}(\bar\sigma^2_{\rm gw})}\right]\,,
\label{e:SSS-reduced-weak-white}
\ee
where 
\be
{\rm var}(\bar\sigma^2_{\rm gw}) \equiv
\frac{1}{N}\bar\sigma^2_{1}\bar\sigma^2_{2}\,,
\label{e:reduced_var_white}
\ee
with
\be
\bar\sigma^2_{1}\equiv \frac{1}{N_{\rm tot}}\sum_{I,i} d_{1,Ii}^2\,,
\qquad
\bar\sigma^2_{2}\equiv \frac{1}{N_{\rm tot}}\sum_{I,i} d_{2,Ii}^2\,
\label{e:bar_sigma_{1,2}}
\ee
being estimates of the total auto-correlated power in the two 
detectors constructed from all the data.
We expect SSC-reduced and SSC-full to perform equally well, assuming $N\gg 1$,
which is needed for the cross-correlation data to be approximately Gaussian.

\subsubsection{SSI-full}

For our proposed stochastic search for intermittent GWBs, we build upon the framework of Drasco and Flanagan \cite{Drasco-Flanagan:2003} and extend their proposed formalism to a larger number of samples per segment ($N\gg 1$) and allow for the amplitudes to be drawn from a uniform-in-volume distribution. The likelihood takes the same form as \eqref{eq:DFmixture}, where the segment-dependent signal and noise likelihoods are now respectively given by:
\begin{align}
&{\cal L}_s(d_I|\langle\sigma_b^2\rangle,\sigma^2_{n_1},\sigma^2_{n_2})
=\int_{\sigma^2_{b,{\rm min}}(\langle\sigma^2_b\rangle)}^{\sigma^2_{b,{\rm max}}}
{\rm d}\sigma^2_{b,I}\>\pi(\sigma^2_{b,I}|\langle\sigma^2_b\rangle)
\frac{1}{(2\pi)^N 
\left(\sigma^2_{1,I} \sigma^2_{2,I} - (\sigma^2_{b,I})^2\right)^{N/2}}
\nonumber\\
&\hspace{2in}\times 
\exp\left\{-\frac{1}{2}
\frac{N}{
\left(\sigma^2_{1,I}\sigma^2_{2,I}-(\sigma^2_{b,I})^2\right)}
\left[\hat \sigma^2_{1,I}\sigma^2_{2,I}
+ \hat \sigma^2_{2,I}\sigma^2_{1,I}
- 2 \hat\sigma^2_{b,I} \sigma^2_{b,I}
\right]\right\}\,,\\
&{\cal L}_n(d_I|\sigma^2_{n_1},\sigma^2_{n_2})
=\frac{1}{(2\pi)^N 
\left(\sigma^2_{n_1} \sigma^2_{n_2}\right)^{N/2}}
\exp\left\{-\frac{N}{2}
\left[\frac{\hat \sigma^2_{1,I}}{\sigma^2_{n_1}}
+ \frac{\hat \sigma^2_{2,I}}{\sigma^2_{n_2}}
\right]\right\}\,,
\end{align}
where
\be
\hat\sigma^2_{b,I} \equiv \frac{1}{N}\sum_{i}d_{1,Ii}d_{2,Ii}\,,
\qquad
\hat\sigma^2_{1,I} \equiv \frac{1}{N}\sum_{i}d^2_{1,Ii}\,,
\qquad
\hat\sigma^2_{2,I} \equiv \frac{1}{N}\sum_{i}d^2_{2,Ii}\,.
\label{e:autopower-reduced_ssi}
\ee
In the above expression for the signal likelihood, we used
\be
\sigma^2_{1,I}\equiv \sigma^2_{n_1} + \sigma^2_{b,I}\,,
\qquad
\sigma^2_{2,I}\equiv \sigma^2_{n_2} + \sigma^2_{b,I}\,,
\ee
which are parameters describing the {\em segment-dependent} 
total auto-correlated power, with the segment dependence
coming from the burst variance $\sigma^2_{b,I}$.

Note that the segment-dependent signal likelihood requires
a marginalization over the segment-dependent burst variances $\sigma^2_{b,I}$, which is taken into account by the appropriate use of prior distribution, as introduced in \eqref{e:p_sigma2b_popavg}:
\be
\pi(\sigma^2_{b,I}|\langle\sigma^2_b\rangle)= 
\frac{\langle\sigma^2_b\rangle (\sigma^2_{b,{\rm max}})^{1/2}}
{\sqrt{-3 + 12{\sigma^2_{b,{\rm max}}}/{\langle\sigma^2_b\rangle}}-3}\,
(\sigma^2_{b,I})^{-5/2}\,,
\label{e:pi_sigma2bI}
\ee
where
\be
\sigma^2_{b,{\rm min}}(\langle\sigma^2_b\rangle) 
= \frac{2 \sigma^2_{b,{\rm max}}}
{6 \sigma^2_{b,{\rm max}}/\langle\sigma^2_b\rangle - 1 - 
\sqrt{-3 + 12{\sigma^2_{b,{\rm max}}}/{\langle\sigma^2_b\rangle}}}\,,
\qquad
\sigma^2_{b,{\rm max}} 
=\sigma^2_{\rm ref}\,\frac{r^2_{\rm ref}}{r^2_{\rm min}}
\label{e:pi_sigma2bI_limits}
\ee
are the limits of integration, which depend on the fixed 
(known) parameter $r_{\rm min}$ and the (unknown) 
population-averaged variance $\langle\sigma^2_b\rangle$.
\subsubsection{SSI-reduced}
\label{s:ssi_red}
Similarly to the case of SSC, one can define a reduced version of the SSI likelihood, provided the number of samples per segment $N$ is large. The segment-dependent signal likelihood still requires
a marginalization over the segment-dependent burst variances
$\sigma^2_{b,I}$:
\be
{\cal L}_s(d_I|\langle\sigma^2_b\rangle,\bar\sigma^2_{n_1},\bar\sigma^2_{n_2})
=\int_{\sigma^2_{b,{\rm min}}(\langle\sigma^2_b\rangle)}^{\sigma^2_{b,{\rm max}}}
{\rm d}\sigma^2_{b,I}\>\pi(\sigma^2_{b,I}|\langle\sigma^2_b\rangle)
\frac{1}{\sqrt{2\pi\,{\rm var}(\bar\sigma^2_{b,I})}}
\exp\left[-\frac{(\hat \sigma^2_{b,I}-\sigma^2_{b,I})^2}{2\,{\rm var}(\bar\sigma^2_{b,I})}\right]\,,
\label{e:SSI-reduced-white-Zs}
\ee
where the prior and limits of 
integration are the same as those used for SSI-full.
In addition, 
\be
{\rm var}(\bar\sigma^2_{b,I}) \equiv
\frac{1}{N}\bar\sigma^2_{1,I}\bar\sigma^2_{2,I}
\label{e:reduced_var_white_unif}
\ee
with
\be
\bar\sigma^2_{1,I}\equiv \bar\sigma^2_{n_1} + \sigma^2_{b,I}\,, 
\qquad
\bar\sigma^2_{1,I}\equiv \bar\sigma^2_{n_2} + \sigma^2_{b,I}\,,
\ee
where we estimate 
the white noise variances from the 
auto-correlated and cross-correlated power in the two 
detector outputs using the full set of data:
\be
\bar\sigma^2_{\rm gw}\equiv \hat\sigma^2_{\rm gw}\theta(\hat\sigma^2_{\rm gw})\,,
\qquad
\bar\sigma^2_{n_1}\equiv 
(\hat\sigma^2_1-\bar\sigma^2_{\rm gw})\theta(\hat\sigma^2_1-\bar\sigma^2_{\rm gw})\,,
\qquad
\bar\sigma^2_{n_2}\equiv 
(\hat\sigma^2_2-\bar\sigma^2_{\rm gw})\theta(\hat\sigma^2_2-\bar\sigma^2_{\rm gw})\,,
\label{e:noisepower-reduced}
\ee
where 
\be
\hat\sigma^2_{\rm gw} \equiv \frac{1}{N_{\rm tot}}\sum_{I,i}d_{1,Ii}d_{2,Ii}\,,
\qquad
\hat\sigma^2_1 \equiv \frac{1}{N_{\rm tot}}\sum_{I,i}d^2_{1,Ii}\,,
\qquad
\hat\sigma^2_2 \equiv \frac{1}{N_{\rm tot}}\sum_{I,i}d^2_{2,Ii}\,.
\label{e:autopower-reduced_ssc}
\ee
In the above expressions,
$\theta(x)$ 
is the usual Heaviside step function, which is defined as 
$\theta(x)=0$ or $1$ depending on whether $x<0$ or $x>0$, and the
hatted quantities $\hat\sigma_{\rm gw}^2$,
$\hat\sigma_{1}^2$, $\hat\sigma_{2}^2$ are the quadratic combinations of the data in the two detectors. This simplification is
possible since the simulated noise is stationary.

The segment-dependent noise likelihood
${\cal L}_n(d_I|\bar\sigma^2_{n_1},\bar\sigma^2_{n_2})$ 
is given as before by:
\begin{equation}
{\cal L}_n(d_I|\bar\sigma^2_{n_1},\bar\sigma^2_{n_2})
= \sqrt{\frac{N}{2\pi\,\bar\sigma^2_{n_1}\bar\sigma^2_{n_2}}}
\exp\left[-\frac{N}{2}\frac{(\hat \sigma^2_{b,I})^2}{\bar\sigma^2_{n_1}\bar\sigma^2_{n_2}}\right]\,.
\end{equation}

\subsection{Likelihoods for colored signals}
\label{app:LikelihoodsColored}

The signal and noise dependent-likelihoods for SSI are specified in Sec. \ref{s:SSI} for both the full (infer noise parameters) and reduced (use estimated noise parameters) analyses. When analyzing stochastic bursts (Sec. \ref{s:Stochastic bursts}) and deterministic chirps (Sec. \ref{s:Deterministic bursts}), the segment prior and integration bounds are specified in \eqref{e:p_Omegab_popavg} and the subsequent paragraph.

\subsubsection{SSC-full}
\label{app:SSSFullColored}
For the continuous search, SSC, the full likelihood is specified by
\begin{align}
&{\cal L}(d|\Omega_{\rm gw},\sigma^2_{n_1},\sigma^2_{n_2})
=\prod_{I=1}^{N_{\rm seg}}
\prod_k
\frac{1}{(\pi T/2)^{2M}(P_{1}(f_k)P_{2}(f_k) - P^2_{\rm gw}(f_k))^{M}}
\nonumber\\
&\hspace{.5in}\times
\exp\left\{-\frac{M}
{(P_{1}(f_k)P_{2}(f_k) - P^2_{\rm gw}(f_k))}\left[
\hat P_{1,Ik}P_{2}(f_k) 
+ \hat P_{2,Ik}P_{1}(f_k)
- 2 \hat P_{{\rm gw},Ik} P_{\rm gw}(f_k)
\right]\right\}\,,
\label{e:SSS_likefinal_colored}
\end{align}
where
\be
P_{1}(f)
\equiv P_{n_1}(f) + P_{\rm gw}(f)\,,
\qquad
P_{2}(f)
\equiv P_{n_2}(f) + P_{\rm gw}(f)\,,
\ee
with
\be
P_{n_1}(f)\equiv
\frac{\sigma^2_{n_1}}{(f_{\rm high} - f_{\rm low})}\,,
\qquad
P_{n_2}(f)\equiv
\frac{\sigma^2_{n_2}}{(f_{\rm high} - f_{\rm low})}\,,
\qquad
P_{\rm gw}(f)\equiv \Omega_{\rm gw} H(f)\,,
\ee
and $H(f)$ is given by \eqref{e:H(f)}. Note that the population parameter for SSC is $\Omega_{\rm gw}$, the time and population-averaged energy density amplitude. In addition, the data enter the signal evidence via the same 
quadratic combinations as for SSI-full (see \eqref{e:SSI-coarse-grained-estimators}),
but with the cross-correlation combination now defining
$\hat P_{{\rm gw},Ik}$ as opposed to $\hat P_{{b},Ik}$.

\subsubsection{SSC-reduced}
\label{app:SSSReducedColored}
For SSC-reduced, we have~\cite{Matas-Romano:2021}:
\begin{align}
{\cal L}(d|\Omega_{\rm gw}, \bar\sigma^2_{n_1}, \bar\sigma^2_{n_2})
&=
\prod_{I=1}^{N_{\rm seg}}
\frac{1}{\sqrt{2\pi\,{\rm var}(\bar\Omega_{\rm gw})}} 
\exp\left[-\frac{(\hat\Omega_{{\rm gw},I}-\Omega_{\rm gw})^2}{2\,{\rm var}(\bar\Omega_{\rm gw})}\right]\,,
\label{e:SSI-reduced-colored-Zs2}
\end{align}
where
\be
\hat \Omega_{{\rm gw},I}
\equiv 
\frac{\sum_k Q(f_k)\hat  P_{{\rm gw},Ik}}
{\sum_{k'} Q(f_{k'}) H(f_{k'})}\,,
\qquad
{\rm var}(\bar\Omega_{\rm gw})
\equiv \left({2M}\sum_k 
Q(f_k)H(f_k)\right)^{-1}
\label{e:optimally-filtered-CC_app}
\ee
are the optimally-filtered cross-correlation estimators and
corresponding variances, which are constructed from coarse-grained estimates of the cross-correlated power
$\hat P_{{\rm gw},Ik}$,
and the optimal filter function
\be
Q(f)
\equiv \frac{H(f)}{\bar P_{1}(f)\bar P_{2}(f)}\,.
\ee
In the above expression,
\be
\bar P_{1}(f)\equiv 
\frac{\bar\sigma^2_{n_1}}{(f_{\rm high} - f_{\rm low})}
+ \bar\Omega_{\rm gw} H(f)\,,
\qquad
\bar P_{2}(f)\equiv
\frac{\bar\sigma^2_{n_2}}{(f_{\rm high} - f_{\rm low})}
+ \bar\Omega_{\rm gw} H(f)\,,
\ee
where $\bar\sigma^2_{n_1}$, $\bar\sigma^2_{n_2}$ 
are measured estimates of the detector noise power
as defined in \eqref{e:noisepower-reduced},
and $\bar\Omega_{\rm gw}$ is related to 
$\bar\sigma^2_{\rm gw}$ (also defined in \eqref{e:noisepower-reduced})
via
\be
\bar\Omega_{\rm gw}=
\frac{4}{3}
\frac{\bar\sigma^2_{\rm gw}}{f_{\rm ref}}
\left(\frac{3H_0^2}{10\pi^2}\frac{1}{f_{\rm ref}^3}\right)^{-1}
\left[
\left(\frac{f_{\rm ref}}{f_{\rm low}}\right)^{4/3}-
\left(\frac{f_{\rm ref}}{f_{\rm high}}\right)^{4/3}
\right]^{-1}\,.
\ee
This last equation follows from the general relation
between variance and power spectrum,
\be
\sigma^2_{\rm gw} 
\equiv \int_{f_{\rm low}}^{f_{\rm high}}{\rm d}f\>P_{\rm gw}(f)
=\Omega_{\rm gw} \int_{f_{\rm low}}^{f_{\rm high}}{\rm d}f\>H(f)
=\Omega_{\rm gw} \left(\frac{3H_0^2}{10\pi^2}\frac{1}{f_{\rm ref}^3}\right)
\int_{f_{\rm low}}^{f_{\rm high}}{\rm d}f\>\left(\frac{f}{f_{\rm ref}}\right)^{-7/3}\,.
\ee

\subsubsection{DSI-full}
\label{app:DSIFullColored}
We also analyze the colored data with DSI, our much simpler implementation of the deterministic-signal-based search.

Following \cite{Smith-Thrane:2018} for two detectors, we define the DSI segment-dependent signal likelihood to be

\be
{\cal L}_s(d_I|r_{\rm max}, \sigma^2_{n_1}, \sigma^2_{n_2}) \propto \int_{r_{\rm min}}^{r_{\rm max}}{\rm d} r_I\ \pi(r_I|r_{\rm max})\exp\left\{-\frac{1}{2}(4\Delta f)\sum_k \sum_{\mu=1,2} \frac{|(\tilde{d}_{\mu,Ik}-\tilde{h}_{\rm chirp}(r_I; f_k))|^2}{P_{n_\mu}}\right\},
\ee
where $\tilde{d}_{\mu,Ik}$ and $\tilde{h}_{\rm chirp}(r_I; f_k)$ are the Fourier transform of the data and chirp waveform, respectively,
with all of the other chirp parameters
assumed to be known a~priori.
In the above signal evidence, we are marginalizing over the 
segment-dependent source distance $r_I$, which is drawn from 
a uniform-in-volume distribution $\pi(r_I|r_{\rm max})$ as given by \eqref{e:unif-in-vol}.

By taking $\tilde{h}_{\rm chirp}(r_I;f_k)=0$ (corresponding to no signal in the data) the corresponding segment-dependent noise likelihood is
\be
{\cal L}_n(d_I|\sigma^2_{n_1}, \sigma^2_{n_2}) \propto \exp\left\{-\frac{1}{2}(4\Delta f)\sum_k\sum_{\mu=1,2} \frac{|\tilde{d}_{\mu,Ik}|^2}{P_{n_{\mu}}}\right\}.
\ee
\subsubsection{DSI-reduced}
\label{app:DSIReducedColored}
For the reduced implementation, we substitute the noise parameters with the auto-correlated power estimates which gives, 
\be
\label{e:DSI_red}
{\cal L}_s(d_I|r_{\rm max}, \bar{\sigma}^2_{n_1}, \bar{\sigma}^2_{n_2}) \propto \int_{r_{\rm min}}^{r_{\rm max}}{\rm d} r_I\> \pi(r_I|r_{\rm max})\exp\left\{-\frac{1}{2}(4\Delta f)\sum_k \sum_{\mu=1,2} \frac{|(\tilde{d}_{\mu,Ik}-\tilde{h}_{\rm chirp}(r_I;f_k))|^2}{\bar P_{n_\mu}}\right\}
\ee
and
\be
{\cal L}_n(d_I|\bar{\sigma}^2_{n_1}, \bar{\sigma}^2_{n_2}) \propto \exp\left\{-\frac{1}{2}(4\Delta f)\sum_k\sum_{\mu=1,2} \frac{|\tilde{d}_{\mu,Ik}|^2}{\bar P_{n_{\mu}}}\right\}
\ee
for the segment-dependent signal and noise likelihoods, respectively.

\bibliography{refs}

\end{document}